\documentclass[apj]{emulateapj}
\slugcomment{{\sc Accepted to ApJ:} Jul 4, 2016}
\usepackage{epsf, graphicx, subfigure}
\usepackage{amsmath}
\usepackage{float}
\usepackage[countmax]{subfloat}
\usepackage[usenames]{color}
\usepackage{hyperref}
\usepackage{natbib}
\usepackage{lipsum}
\usepackage{multirow}

\renewcommand{\vec}[1]{\boldsymbol{#1}}

\shorttitle{Broad Emission Line Profiles from SBHBs}
\shortauthors{Nguyen and Bogdanovi\'c.}

\begin{document}

%%% These are Mike's colored text macros (first red then blue). This
%%% first definition in each pair disables the macro while the second
%%% enables it. To disable, comment out the second defintiion of each
%%% pair. 

\def\redtext#1{#1}
\def\redtext#1{{\color{red}#1}}
\def\bluetext#1{#1}
\def\bluetext#1{{\color{blue}#1}}
\def\greentext#1{#1}
\def\greentext#1{{\color{green}#1}}
%\def\ergs{\ifmmode{~{\rm erg~s^{-1}}}\else{~erg~s$^{-1}$}\fi}
%%% This is an ionic species macro, adapted from AASTeX
%\newcounter{species} 
%\def\ion#1#2{\setcounter{species}{#2}#1$\;${\scriptsize\Roman{species}}\relax}

%-------------------------------------------------------------------

\title{Emission Signatures from Sub-parsec Binary Supermassive Black Holes I: Diagnostic Power of Broad Emission Lines}
%\title{Emission Signatures from Sub-parsec Binary Supermassive Black Holes II: Effect of Accretion Disk Wind on Broad Emission Lines}
%\title{Emission Signatures from Sub-parsec Binary Supermassive Black Holes III: Comparison of Models with Observations}

\author{Khai Nguyen\altaffilmark{1} and Tamara Bogdanovi\'c\altaffilmark{1}}
\altaffiltext{1}{Center for Relativistic Astrophysics, School of Physics, Georgia Institute of Technology, Atlanta GA 30332, USA }

\begin{abstract}
Motivated by advances in observational searches for sub-parsec supermassive black hole binaries (SBHBs) made in the past few years we develop a semi-analytic model to describe spectral emission line signatures of these systems. The goal of this study is to aid the interpretation of spectroscopic searches for binaries and help test one of the leading models of binary accretion flows in the literature: SBHB in a circumbinary disk. In this work we present the methodology and a comparison of the preliminary model with the data. We model SBHB accretion flows as a set of three accretion disks: two mini-disks that are gravitationally bound to the individual black holes and a circumbinary disk. Given a physically motivated parameter space occupied by sub-parsec SBHBs, we calculate a synthetic database of nearly 15 million broad optical emission line profiles and explore the dependence of the profile shapes on characteristic properties of SBHBs. We find that the modeled profiles show distinct statistical properties as a function of the semi-major axis, mass ratio, eccentricity of the binary, and the degree of alignment of the triple disk system. This suggests that the broad emission line profiles from SBHB systems can in principle be used to infer the distribution of these parameters and as such merit further investigation. Calculated profiles are more morphologically heterogeneous than the broad emission lines in observed SBHB candidates and we discuss improved treatment of radiative transfer effects which will allow direct statistical comparison of the two groups.
\end{abstract}

           
\section{Introduction}\label{sec:intro}

The past ten years have marked a period of active research on supermassive black hole (SBH) pairs and binaries spearheaded by theoretical studies which have investigated how black holes grow, form pairs and interact with their environment. Interest in them has been driven by a realization that SBHs play an important role in evolution of their host galaxies \citep{ferrarese00,gebhardt00,tremaine02} and also, by intention to understand the parent population of merging binaries which are the prime targets for the long anticipated space-based gravitational wave (GW) observatories. We refer to dual SBHs at large separations as {\it pairs} and to the gravitationally bound SBHs as {\it binaries}, hereafter.

Theoretical studies have established that evolution of SBH pairs from kiloparsec to smaller scales is determined by gravitational interactions of individual black holes with their environment \citep{bbr80,mayer13}. These include interaction of the SBHs with their own wakes of stars and gas \citep[a.k.a., dynamical friction;][]{chandra43, ostriker99,mm01,escala04} and scattering of the SBHs by massive gas clouds and spiral arms produced by local and global dynamical instabilities during the merger \citep{fiacconi13,roskar14}. During these interactions the SBHs exchange orbital energy and angular momentum with the ambient medium and can in principle grow though accretion  \citep{escala04, escala05, dotti06, dotti07, dotti09, callegari11, khan12, chapon13}. These factors determine the SBH dynamics and whether they evolve to smaller separations to form a gravitationally bound binary. For example, \citet{callegari09,callegari11} find that SBH pairs with mass ratios $q<0.1$ are unlikely to form binaries within a Hubble time at any redshift. On the other hand SBH pairs with initially unequal masses can evolve to be more equal-mass, through preferential accretion onto a smaller SBH. It is therefore likely that SBH pairs with $q\gtrsim0.1$ form a parent population of bound binaries at smaller separations.

Gravitationally bound binary forms at the point when the amount of gas and stars enclosed within its orbit becomes comparable to the total mass of the two black holes.  For a wide range of host properties and SBH masses this happens at orbital separations $\lesssim 10$~pc \citep{mayer07,dotti07,khan12}. The subsequent rate of binary orbital evolution depends on the nature of gravitational interactions that it experiences and is still an area of active research often abbreviated as  {\it the last parsec problem}.  The name refers to a possible slow-down in the orbital evolution of the parsec-scale supermassive black hole binaries (SBHBs) caused by inefficient interactions with stars \citep{mm01} and gas \citep{escala05}. If present, a consequence of this effect would be that a significant fraction of SBHBs in the universe should reside at orbital separations of $\sim 1$pc. Several recent theoretical studies that focus on the evolution of binaries in predominantly stellar backgrounds however report that evolution of binaries to much smaller scales continues unhindered \citep{berczik06, preto11, khan11,khan12a,khan13}, although the agreement about the leading physical mechanism responsible for the evolution is still not universal \citep{vasiliev14}. 

SBH binaries in predominantly gaseous environments have also been a topic of a number of theoretical studies \citep{an05, macfadyen08, bogdanovic08, bogdanovic11, cuadra09, haiman09, hayasaki09, roedig12,shi12,noble12,kocsis12b, kocsis12a,dorazio13,farris14,rafikov16}. They find that binary torques can truncate sufficiently cold circumbinary disks and create an inner low density cavity by evacuating the gas from the central portion of the disk \citep[see][and references above]{lp79}. SBHs in this phase can accrete by capturing gas from the inner rim of the circumbinary disk and can in this way maintain mini-disks bound to individual holes. As the binary orbit decays, the inner rim of the circumbinary disk follows it inward until the timescale for orbital decay by gravitational radiation becomes shorter than the viscous timescale\footnote{The time scale on which the angular momentum is transported outwards through the disk.} of the disk \citep{an05}. At that point, the rapid loss of orbital energy and angular momentum through gravitational radiation cause the binary to detach from the circumbinary disk and to accelerate towards coalescence. 

Through its dependence on the viscous time scale, orbital evolution of a gravitationally bound SBHB in the circumbinary disk depends on thermodynamic properties of the disk. These are uncertain, as they are still prohibitively computationally expensive to model from first principles and are unconstrained by observations. More specifically, the thermodynamics of the disk is determined by the binary dynamics and also the presence of magnetic field and radiative heating and cooling of the gas. While the role of magnetic field in circumbinary disks has been explored in some simulations \citep{giaco12,shi12,noble12,farris12,gold14}, a fully consistent calculation of radiative processes is still beyond computational reach. Consequently, current theoretical models can be formulated as parameter studies, where difficult-to-model processes are parametrized in some fashion, but cannot uniquely predict the properties of the circumbinary regions or the emission signatures of SBHBs. The circumbinary disk model is therefore an appealing theoretical concept that must be tested through observations.

Along similar lines, observations of the orbital properties of SBHBs are key to understanding binary evolution. This is because the frequency of binaries as a function of their orbital separation is directly related to the rate at which binaries evolve towards coalescence. Theoretical models predict that the exchange of angular momentum with the ambient medium is likely to result in SBHB orbits with eccentricities $\gtrsim 0.1$, with the exact value depending on whether gravitationally bound SBHs evolve in mostly stellar or gas rich environments \citep{roedig11,sesana11,khb15}. Known semi-major axis and eccentricity distributions would therefore provide a direct test for a large body of theoretical models. 

Our understanding of spin magnitudes and orientations in binary SBHs also relies on theoretical considerations. Interest in this topic was triggered by the prediction of numerical relativity that coalescence of SBHs with certain spin configurations can lead to the ejection of a newly formed SBH from its host galaxy. This effect arises due to the asymmetry in emission of GWs in the final stages of an SBH merger and can lead to a GW kick of up to $\sim5000\,{\rm km\,s^{-1}}$ \citep{campanelli07,lousto11}. Several subsequent theoretical studies found  that accretion and gravitational torques can act to align the spin axes of SBHs evolving in gas rich environments and in such way minimize the GW recoil  \citep{bogdanovic07,dotti10,dotti13,sorathia13,miller13}\footnote{See however \citet{lodato13} for a different view.}.  Mutual SBH spin alignment is  not expected in gas poor environments, geometrically thick, turbulent and magnetically dominated disks \citep{fragile05,fragile07,mckinney13}, allowing a possibility that runaway SBHs and empty nest galaxies may exist. Hence, if observations can independently provide an insight into the geometry of circumbinary disks and spin properties of SBHBs, they would be an important probe of the alignment hypothesis. 

In this work, which constitutes part I of a series, we develop a preliminary semi-analytic model of an SBHB in circumbinary disk and use it to calculate a database of broad, optical emission line profiles associated with binary systems. Our analysis indicates that such profiles can in principle be used to infer statistical distribution of SBHB parameters, making modeling of binary spectroscopic signatures a worthwhile task. The comparison of the preliminary model with the data from observed SBHB candidates however indicates that further improvements to the model are necessary before the synthetic profiles can be used to interpret the observations. These steps will be presented in subsequent papers.

This paper is organized as follows: Section~\ref{sec:observations} describes the status of ongoing spectroscopic searches for SBHBs, Section~\ref{sec:calculation} and the Appendix outline the semi-analytic model used in calculation of SBHB emission line signatures, and Section~\ref{sec:results} gives the description of results. We discuss the validity of assumptions and implications of our results in Section~\ref{sec:discussion} and conclude in Section~\ref{sec:conclusions}.

%====================================================================================================%

\section{Status of Spectroscopic Searches for SBHBs}\label{sec:observations}

Key characteristic of gravitationally bound SBHBs is that they are observationally elusive and expected to be intrinsically rare. Theorists estimate that a fraction $<10^{-3}$ of active galactic nuclei (AGNs) at redshift $z<0.7$ may host SBHBs \citep{volonteri09}. This result implies that any observational search for SBHBs must involve a large sample of AGNs and that observational technique used in the search needs to distinguish signatures of binaries from those of AGNs powered by single SBHs. 

Spectroscopic searches rely on the detection of the Doppler-shift in the emission line spectrum of an SBHB candidate that arise as a consequence of the binary orbital motion. This approach is reminiscent of a well established technique for detection of the single- and double-line spectroscopic binary stars. In both classes of spectroscopic binaries, the lines are expected to oscillate about their local rest frame wavelength on the orbital time scale of a system.  In the context of the binary model, the spectral emission lines are assumed to be associated with the gas accretion disks that are gravitationally bound to the individual SBHs \citep{gaskell83,gaskell96,bogdanovic08}.  Given the velocity of the bound gas the emission line profiles from the SBH mini-disks are expected to be Doppler-broadened, similar to the emission lines originating in the broad line regions (BLRs) of AGNs. Moreover, several theoretical studies have shown that in unequal mass binaries accretion occurs preferentially onto the lower mass object \citep{al96,gr00,hayasaki07}, rendering it potentially more luminous than the primary. If so, this indicates that some fraction of SBHBs may appear as the single-line spectroscopic binaries.

This realization lead to a discovery of a number of SBHB candidates based on the criterion that the culprit sources exhibit broad optical lines offset with respect to the rest frame of the host galaxy \citep{bogdanovic09a,dotti09a,bl09,tang09,decarli10, barrows11,tsal11, tsai13}\footnote{In an alternative approach anomalous line ratios have been used to flag SBHB candidates with perturbed BLRs \citep{montuori11,montuori12}.}.
 Because this effect is also expected to arise in the case of a recoiling SBH receding from its host galaxy, the same approach has been used to flag candidates of that type \citep{komossa08a,shields09,civano10, robinson10,lusso14}. The key advantage of the method is its simplicity, as the spectra that exhibit emission lines shifted relative to the galaxy rest frame are relatively straightforward to select from large archival data sets, such as the Sloan Digital Sky Survey (SDSS). Its main complication however is that the Doppler shift signature is not unique to these two physical scenarios and complementary observations are needed in order to determine the nature of the observed candidates \citep[e.g.,][]{popovic12, bogdanovic14}.

To address this ambiguity a new generation of spectroscopic searches has been designed to monitor the offset of the broad emission line profiles over multiple epochs and target sources in which modulations in the offset are consistent with the binary orbital motion \citep{eracleous12,bon12,decarli13,shen13,ju13, liu13, li16}. For example, \citet{eracleous12} searched for $z<0.7$ SDSS quasars whose broad $H\beta$ lines are offset by $\gtrsim 1000\;{\rm km\,s^{-1}}$ and selected 88 quasars for observational followup from the initial group of $\sim 16,000$ objects. After the second and third epoch of observations of this sample, statistically significant changes in the velocity offset were found in 14 \citep{eracleous12} and 9 objects \citep{mathes14}, respectively, in broad agreement with theoretical predictions for frequency of SBHBs \citep{volonteri09}.

%====================================================================================================%

\section{Description of the Model}\label{sec:calculation}

\subsection{Emission line profiles from SBHB in circumbinary disk}\label{sec:cbd}  

Motivated by theoretical models described in the literature and ongoing observations we consider the sub-parsec binaries with mass ratios in the range $0.1\leq q \leq 1$, where $q=M_2/M_1$. The orbital separation and period of such binaries can be expressed in terms of the spectroscopically determined velocity offset, which is their key observable property. If the measured velocity offset can be attributed to the motion of the secondary SBH, as indicated by the accretion rate inversion found in theoretical studies of SBHBs (see Section~\ref{sec:observations}), then the projected velocity of the secondary, $u_2$, is related to its true orbital speed, $v_{\rm orb2}$, as $u_2$ = $v_{\rm orb2}\,\sin{i}\,|\sin{\phi}|$. Here $i$ is the inclination of the orbital axis of the binary relative to observer's the line of sight ($i=0$ is face-on) and $\phi$ is the orbital phase at the time of the observation ($\phi = 0$ corresponds to conjunction). Note that the expression for $u_2$ applies to circular orbits, an assumption which we use to obtain illustrative estimates but in our model calculations actually consider both circular and eccentric orbits. Following \citet{eracleous12} we express the period and orbital separation in terms of the total mass $M_8 = (M_1 + M_2)/10^8\,M_\odot$ and the projected velocity of the secondary, $u_{2,3} = u_2/10^3\,{\rm km\,s^{-1}}$ as
\begin{eqnarray}
a=\frac{0.11\,M_8}{(1+q)^2\,u_{2,3}^2}
\left(\frac{\sin{i}}{\sin{45^\circ}} \frac{|\sin{\phi}|}{\sin{45^\circ}}\right)^2\,{\rm pc}\\
P=\frac{332\,M_8}{(1+q)^3\,u_{2,3}^3}
\left(\frac{\sin{i}}{\sin{45^\circ}} \frac{|\sin{\phi}|}{\sin{45^\circ}}\right)^3\,{\rm yr}
\end{eqnarray}
If the measured velocity offset is instead associated with the primary SBH, the above expressions can be written in terms of the projected velocity of the primary, $u_{1,3} = q\,u_{2,3}$ where $u_{1,3} = u_1/10^3\,{\rm km\,s^{-1}}$. In the expressions above we choose $i=\phi=45^{\circ}$ and discuss the parameter values used in our model calculations in Section~\ref{sec:params}.

The accretion flow is described as a set of three circular accretion disks: two mini-disks that are gravitationally bound to the individual SBHs and a circumbinary disk. The three disks are modeled as independent BLRs, where the extent of the two mini-disks, as well as the central opening in the circumbinary disk are constrained by the size of the binary orbit and are subject to tidal truncation by the binary SBH \citep{paczynski77,lp79,sepinsky07}. In this model both accreting SBHs can shine as AGNs and illuminate their own mini-disk as well as the two other disks in the system. We assume that the bolometric luminosity of each AGN correlates with the accretion rate onto its SBH and that photoionization by the AGNs gives rise to the broad, low-ionization optical emission lines just like in ``ordinary" BLRs \citep{csd89,csd90}. The emissivity of each disk can then be evaluated as a function of the accretion rate onto the SBHs and the disk size. We utilize the published measurements of accretion rates from simulations of SBHBs \citep{al96,gr00,hayasaki07,roedig11,farris14} in order to establish the relative bolometric luminosities of the two AGN in a binary and emissivity of each disk component. Any assumptions about the mutual orientation of the two mini-disks and circumbinary disk are relaxed and they are allowed to assume arbitrary orientations relative to the observer.

\begin{deluxetable}{cl}  % <--- column justification (center/left/right)
\tabletypesize{\scriptsize}
\tablecolumns{4}
%\tablewidth{300pt}
\tablewidth{0pt} 
\tablecaption{Parameters of the model.}\label{table:parameters}
\tablehead{   % column headings
Parameter & Value}
\startdata
$q$ & 1 , 9/11 , 2/3 , 3/7 , 1/3 , 1/10 \\
$a/M$ & $5\times 10^3$ , $10^4$ , $5 \times 10^4$, $10^5$, $10^6$ \\
$e$ & 0.0 , 0.5 \\
$f$ & $0^{\circ}$, $72^{\circ}$, $144^{\circ}$, $216^{\circ}$, $288^{\circ}$ \\ 
$R_{\rm in1}/M_1$, $R_{\rm in2}/M_2$ & 500 , 1000\\
$R_{\rm out3}$ & $3a$ \\
$i$ & $5^{\circ}$, $55^{\circ}$, $105^{\circ}$, $155^{\circ}$\\
$\phi$  & $0^{\circ}$, $36^{\circ}$, $108^{\circ}$, $180^{\circ}$, $242^{\circ}$, $324^{\circ}$\\ 
$\theta_1$, $\theta_2$ & $0^{\circ}$, $30^{\circ}$, $60^{\circ}, 105^{\circ}$, $135^{\circ}$, $165^{\circ} $\\
$\phi_1$, $\phi_2$ & $0^{\circ}$, $25^{\circ}$, $60^{\circ}$, $185^{\circ}$, $210^{\circ}$, $235^{\circ}$\\
$h_1/M_1$, $h_2/M_2$ & 10\\
$\sigma/{\rm km\,s^{-1}}$ & 850
\enddata
\tablecomments{See text for description of parameters.}
\label{table:parameters}
\end{deluxetable}

 We follow the line profile calculations described in \citet{chen89}, \citet{chen89b} and \citet{eracleous95} to obtain an emission line profile from each Keplerian, circular, relativistic thin disk in the weak-field approximation. Such line profiles are characteristic of rotating disks and resemble the persistent, double-peaked Balmer lines found in about 10 -- 20\% of broad-line radio galaxies and about 3\% of all active galaxies \citep{eh94,eh03,strateva03}. More generally, several works have demonstrated that disk models of this type can be used to describe emission from BLRs of most AGNs when additional radiative transfer effects of the disk atmosphere on the emission line profiles are accounted for \citep{cm96, mc97, flohic12, chajet13}. We adopt this approach because of its broad applicability, considering the baseline model first and addressing the mentioned radiative transfer effects in Paper~II of the series (see also Section~\ref{secsec:approx} for more discussion).
 
 The emission line fluxes from the three disks are evaluated and summed following the steps presented in the Appendices~\ref{sec:appendixsingle} and  \ref{appendixbinary}.  The main objective of this calculation is to obtain the final expression for the flux density in the observer's frame as a function of parameters defined in the reference frame of the binary. Using this approach we calculate a database of profiles by drawing from a parameter space that describes different configurations of SBHBs and their associated circumbinary regions. For a somewhat different approach see also the work by \citet{simic16} who model the SBHB accretion flow as either one or two BLRs, each of which contributes a Gaussian emission line profile.

In this work we focus on the H$\beta$ emission line profiles, the second line of the hydrogen Balmer series, but note that this calculation is applicable to all permitted, low-ionization broad emission line profiles. The broad emission lines of particular interest are H$\alpha~\lambda6563$, H$\beta\;\lambda4861$, and Mg II~$\;\lambda2798$ because they are prominent and relatively easy to identify in AGN spectra.  These low-ionization lines have been established as reliable tracers of dense gas in BLRs and are presently used in spectroscopic searches for SBHBs at low (H$\alpha$, z $<$ 0.4) and high redshift (Mg\,II, z $<$ 2.5).

\begin{figure}[t]
\centering
\includegraphics[width=.4\textwidth, clip=true]{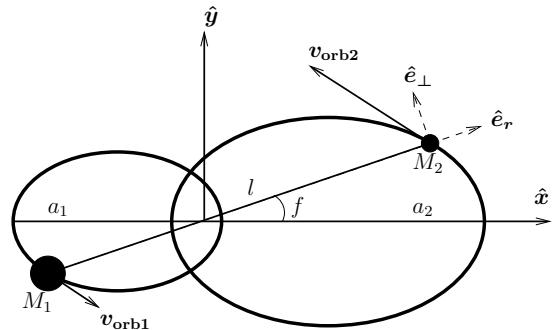}
\caption{Illustration of SBHB geometry for face-on orientation of the orbit. SBHB center of mass marks the origin of the coordinate system which $z$-axis points in the direction of the orbital angular momentum of the binary (out of the page). The $x$-axis points towards the pericenter of the primary SBH orbit and is parallel to the orbital semi-major axis of the binary, $a=(a_1+a_2)$. The orbital phase $f$ is measured counter-clockwise from the $x$-axis to the instantaneous location of secondary SBH. The mini-disks and circumbinary disk are not shown. See text for definition of other variables.}
\label{fig:binary}
\end{figure}

\subsection{Parameters of the model}\label{sec:params}

Table~\ref{table:parameters} summarizes parameter choices for the modeled configurations of SBHB systems. The parameters encode the intrinsic properties of the binary, such as the orbital semi-major axis, eccentricity, the alignment of the triple disk system, as well as the orientation of the SBHB with respect to the distant observer's line of sight. The sample includes 2,545,200 realizations of binaries on circular orbits and 12,273,000 on elliptical orbits, for a total of 14,818,200 configurations. We describe our parameter choices below and present the details of profile calculation in the Appendices.

\begin{itemize}
\item \textit{SBHB mass ratio, $q$} -- Simulations of galaxy mergers that follow pairing of their massive black holes find that SBH pairs with mass ratios $q<0.1$ are unlikely to form gravitationally bound binaries within a Hubble time at any redshift \citep{callegari09,callegari11}. They also find that SBH pairs with initially unequal masses can evolve to be more equal-mass, through preferential accretion onto a smaller SBH. Motivated by these results we choose six values of $q$ in the range $0.1-1$ to represent the mass ratio of the binary.
\item \textit{Semi-major axis, $a$} --  To describe orbital separations of gravitationally bound binaries we chose five values of $a$ ranging from $5000\,M$ to $10^6\, M$, where we use the mass of the binary $M\equiv GM/c^2 = 1.48\times10^{13}\,{\rm cm}\, (M/10^8\,M_\odot)$ as a measure of length in geometric units, where $G=c=1$. For example, for the total mass of the binary of $10^8 M_{\odot}$ this range of semi-major axes corresponds to binary separations $\sim 0.02 - 5$\,pc.  
\item \textit{Orbital eccentricity, $e$} -- Theoretical models that follow evolution of the orbital eccentricity of SBHBs in circumbinary disks suggest that the exchange of angular momentum between them drives a steady increase in binary eccentricity which saturates in the range $0.6-0.8$ \citep{an05,cuadra09,roedig11}. For the purposes of this calculation we choose two values of eccentricity, $e=0.0$ and 0.5, to model SBHBs on both circular and elliptical orbits. 
\item \textit{Orbital phase, $f$} -- Five values of the orbital phase are chosen to describe orbital evolution of SBHBs. $f$ is measured from the positive $x$-axis to the instantaneous location of the secondary SBH in counter-clockwise direction, as illustrated in Figure~\ref{fig:binary}.

\begin{figure}[t]
\centering
\includegraphics[width=.42\textwidth, clip=true]{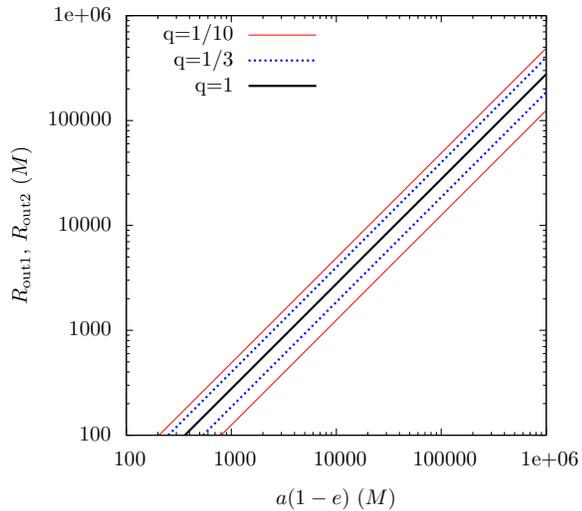}
\caption{Outer radii of the primary and secondary mini-disks as a function of $a$ and $e$. Lines mark SBHB mass ratios $q = 1/10$ (thin, red lines) , 1/3 (dotted, blue) and 1 (solid, black) based on the model of \citet{paczynski77}. For unequal mass ratios the top line marks the size of the larger (primary) mini-disk.}
\label{fig:Rout}
\end{figure}

\item \textit{SBHB accretion rate ratio, $\dot{m}$} -- In the context of this model we assume that the emissivity of each broad emission line region is a function of the AGN luminosity and the disk size. In order to establish the relative bolometric luminosities of the two AGN we compile from the literature the values of $\dot{m} = \dot{M}_2/\dot{M}_1$ and parametrize it as a linear function of $q$ for SBHBs  on circular and eccentric orbits. 
\begin{align}\label{eq:mscale}
\dot{m}\simeq 
\begin{cases}
5.5-4.5q & e=0.0 \\
1.5-0.5q & e=0.5
\end{cases}
\end{align}
Here $\dot{M}_1$ and $\dot{M}_2$ are the accretion rates onto the primary and secondary SBH, respectively. The two relations capture two key results observed in hydrodynamic simulations of prograde SBHBs  \citep[rotating in the same sense as the circumbinary disk;][]{hayasaki07, roedig11}: (1) in unequal mass binaries accretion occurs preferentially onto the smaller of the SBHs and (2) the inversion of accretion rates is more severe for SBHBs on circular orbits. This trend has also been captured by other simulations \citep{al96,farris14} and models motivated by them \citep{gerosa15,young15}.
\item \textit{Size of the broad line regions, $R_{\rm {in}\it{i}}$ and $R_{\rm {out}\it{i}}$} --  Each disk in the triple disk system has an associated BLR defined by a pair of inner and outer radii. In the case of the mini-disks we choose two different values for the BLR inner radius $R_{\rm {in}\it{i}}=500M_{i}$ and $1000M_{i}$, where $i=1,2$ mark the BLR around the primary and secondary SBH, respectively. These choices are motivated by characteristic values for the inner radius of the BLRs in AGNs powered by single SBHs, which emission lines are well modeled by the emission from a Keplerian disk \citep[for e.g.,][]{eh94, eh03}. The outer radii are naturally determined by the tidal torques of the binary and do not extend beyond the Roche lobes of their SBHs. We follow the approach described by \citet{paczynski77} to estimate the average values of $R_{\rm out1}$ and $R_{\rm out2}$ based on the binary separation $a$, and mass ratio $q$ (see Figure~\ref{fig:Rout}). 

We define the size of the circumbinary disk BLR in terms of the SBHB semi-major axis, $R_{\rm in3}=2a$ and $R_{\rm out3}=3a$. The value of the inner radius is directly motivated by theory and simulations which show that SBHB torques create a low density hole with radius about $2a$ in the center of the circumbinary disk \citep{lp79, an05, macfadyen08}. The value of the outer radius of the circumbinary disk BLR is poorly constrained and for the purposes of this calculation we adopt $R_{\rm out3}=3a$. Note that the BLR sizes assumed in this work are consistent with the plausible range empirically derived for low redshift AGN by \citet{kaspi05}.
\item \textit{Emissivity of the broad line regions, $\epsilon_{i}$} -- Each disk in the system is further characterized by the emissivity of the BLR, which arises due to the illumination by the two AGNs. For example, the emissivity of the mini-disk around the primary SBH can be expressed as $\epsilon_{1} = \epsilon_{11} + \epsilon_{12}$, where $\epsilon_{11}$ and $\epsilon_{12}$ correspond to the components of emissivity due to the illumination by its own AGN and the AGN associated with the secondary SBH, respectively. The emissivity of each mini disk associated with its own AGN is described as a power law in radius, with the power law index $p=3$, such that $\epsilon_{11} \propto \epsilon_{22} \propto R^{-p}$ \citep{csd89}. The component of emissivity associated with the companion AGN ($\epsilon_{12}$ and $\epsilon_{21}$) is calculated as a function of its distance and orientation of the mini-disk (Appendix~\ref{appendixbinary}). The emissivity of the circumbinary disk, $\epsilon_{3}$, is calculated as a sum of emissivities due to the two off-center AGN associated with the primary and secondary SBHs. 
\item \textit{Orientation of the observer relative to the binary orbit, $i$ and $\phi$} -- We choose four values of the inclination angle, $i$, to describe the orientation of the observer's line of sight relative to the vector of orbital angular momentum of the SBHB. For example, $i=0^\circ$ represents the clockwise binary seen face-on and values $i>90^\circ$ represent counter-clockwise binaries. Furthermore, we select six values of the azimuthal angle $\phi$ measured in the binary orbital plane, from the positive $x$-axis to the projection of the observer's line of sight, in counter-clockwise direction. For circular SBHBs varying the true anomaly $f$ is equivalent to varying the azimuthal orientation of the observer and in this case we adopt a single nominal value of $\phi=0^\circ$ in calculation of the emission line profiles. However, in the case of eccentric SBHBs we explore a full range of $f$ and $\phi$ angles.
\item \textit{Orientation of the mini disks, $\theta_i$ and $\phi_i$} -- We relax assumptions about the orientation of the mini disks with respect to the binary orbit in order to study how profile shapes depend on it. We choose six values of the polar angle ($\theta_i$) and azimuthal angle ($\phi_i$) to describe the orientation of each mini disk with respect to the vector of orbital angular momentum of the binary. For example, when $\theta_1=\theta_2=0^{\circ}$, both mini-disks are coplanar with the SBHB orbit and for $\theta_i>90^\circ$, the gas in the mini-disks exhibits retrograde motion relative to the circumbinary disk. The azimuthal angles $\phi_i$ are measured in the binary orbital plane, from the positive $x$-axis to the projection of the rotation axis of the mini-disk, in counter-clockwise direction. The circumbinary disk is assumed to always be coplanar and in co-rotation with the binary orbit.
\end{itemize}

\begin{figure}[t]
\centering
\includegraphics[width=0.45\textwidth, clip=true]{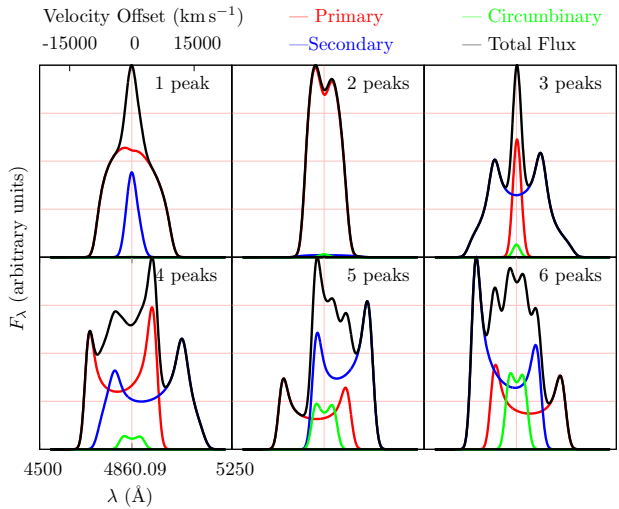}
\caption{Illustration of profile shapes represented in the emission line database. Total flux (black line) is a sum of components contributed by the primary (red), secondary (blue) and circumbinary disk (green). Flux is shown in arbitrary units against wavelength (bottom $x$-axis) and corresponding velocity offset relative to the binary center of mass (top $x$-axis). Pink vertical line at $4860.09 \AA$ marks the rest wavelength of the H$\beta$ emission line.}
\label{fig:diversity}
\end{figure}

Note that some of the model parameters described above are actually not free parameters, because they are constrained by the relevant physical processes and can be expressed in terms of the properties of the binary (this is the case with $\dot{m}$, $R_{\rm out1}$,  $R_{\rm out2}$ and $R_{\rm in3}$). The calculation of the emission line profiles requires definition of two additional parameters which have a lesser impact on their shape. Specifically, motivated by the X-ray studies of the broad iron line reverberation \citep[see review by][]{uttley14}, we assume that the central source of the continuum radiation associated with each SBH is compact and has spatial extent of $h_i=10M_i$. Similarly, we describe the broadening of the emission line profiles due to the random (turbulent) motion of the gas in each disk as $\sigma=850\,{\rm km\,s^{-1}}$.  We discuss implications of our parameter choices in Section~\ref{sec:discussion}.

%====================================================================================================%

\section{Results}\label{sec:results}

In this section we  draw attention to unique features of the modeled population of profiles (Section~\ref{sec:features}) and characterize their shapes in terms of commonly used statistical distribution functions (Section~\ref{sec:stats}). We then investigate whether the complex, composite profile shapes preserve any dependence on the parameters of the underlying SBHB model (Section~\ref{sec:dependence}). 

\begin{deluxetable}{cccccccc} % <--- column justification (center/left/right)
\tablecolumns{4}
\tablecaption{Number of peaks.}
\tabletypesize{\scriptsize}
\tablewidth{0pt}
\tablehead{   % column headings
 & $a$ &1  & 2  & 3  & 4  & 5  & 6 \\
& $(M)$ & (\%) & (\%) & (\%) & (\%) & (\%) &  (\%)
}
\startdata
%\multicolumn{8}{ c }{Circular Orbit $(e=0.0)$} \\
\multirow{5}{*}{\rotatebox{90}{Circular}} & $5\times10^3 $ & 15.01 & 40.15 & 35.84 & 8.94 & 0.06 & 0.00 \\
& $10^4 $   & 28.16 & 54.34 & 17.05 & 0.45 & 0.00 & 0.00 \\
& $5\times10^4 $ & 77.25 & 22.74 & 0.01 & 0.00 & 0.00 & 0.00 \\
& $10^5 $   & 87.07 & 12.93 & 0.00 & 0.00 & 0.00 & 0.00 \\
& $10^6 $   & 100.00 & 0.00 & 0.00 & 0.00 & 0.00 & 0.00\\
\hline
%\multicolumn{8}{ c }{Eccentric Orbit $(e=0.5)$} \\
\multirow{5}{*}{\rotatebox[origin=c]{90}{Eccentric}} & $5\times10^3$ & 5.85 & 14.21 & 50.05 & 22.82 & 6.51 & 0.57 \\
& $10^4$   & 9.03 & 42.36 & 44.05 & 4.39 & 0.17 & 0.00 \\
& $5\times10^4$ & 37.84 & 61.75 & 0.41 & 0.00 & 0.00 & 0.00 \\
& $10^5 $   & 63.97 & 35.98 & 0.05 & 0.00 & 0.00 & 0.00 \\
& $10^6 $   & 100.00 & 0.00 & 0.00 & 0.00 & 0.00 & 0.00
\enddata
%\tablenotetext{*}{100\% of profiles with $a=10^6 M$ is single peaked.}
\label{tab:peaks}
\end{deluxetable}

\begin{figure*}[t]
\centering
\includegraphics[width=0.75\textwidth, clip=true]{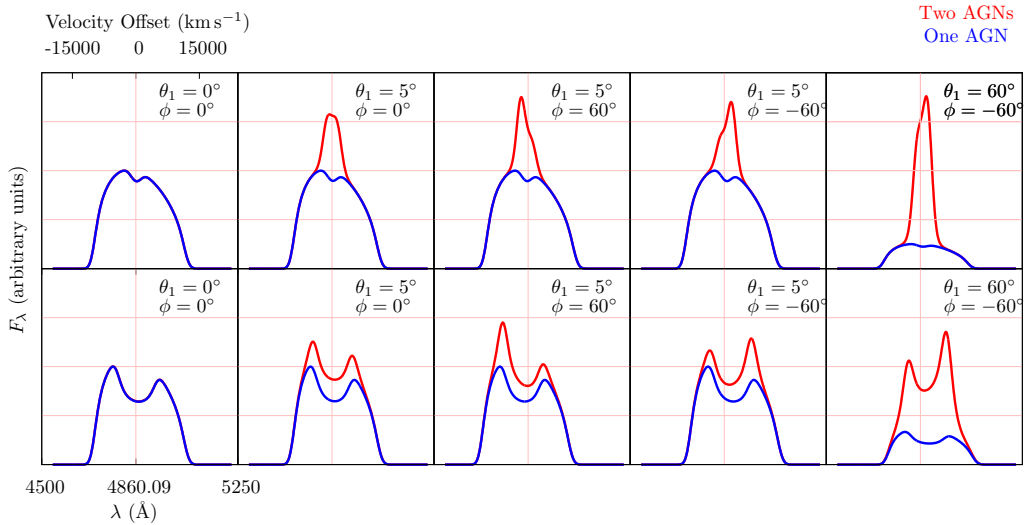}
\caption{Profiles emitted by a single (primary) mini-disk illuminated by its central AGN only (blue line) and by the companion AGN (red). Contribution to the total line flux from illumination by the secondary AGN is negligible only when the mini-disk and the binary orbit are close to coplanar ($\theta_1 = 0^{\circ}$). Excess flux appears in the blue (red) wing of the emission line for the hotspot that moves toward (away from) the observer, as indicated by the azimuthal angle $\phi$. Flux scaling is the same in all panels except for the last one, where profile flux was divided by a factor of 4 for visualization purposes. The sequence of profiles at the top and bottom are calculated for two arbitrary SBHB orbital configurations with $a=50000\,M$  and $5000\,M$, respectively. Parameters shared by both systems are $q=1$, $e=0$, $i=55^{\circ}$, $\phi_1=0^{\circ}$, $R_{\rm in1}=500\,M_1$.}
\label{fig:mutual}
\end{figure*}

\subsection{Characteristic features of the modeled emission line profiles}\label{sec:features}

The most striking property of the modeled emission line profiles is that they can have multiple peaks and their appearance can vary significantly over time, due to the orbital motion of the binary and the resulting variable illumination of the three disks by the two AGNs. Each disk in the triple disk system can give rise to either a single-peaked or a double-peaked profile, depending on the size of its emission region and its orientation with respect to the observer's line of sight. Generally, the larger the BLR, the more likely is the disk to produce a single-peaked profile. This is because the bulk of the emission is contributed by the outer regions on the disk characterized by lower rotational velocities. Similarly, the lower the inclination of the disk with respect to the observer, the more likely it is that the observed profile is single-peaked since the gas velocity along the line of sight is low. Since in our model we account for a range of BLR sizes and inclinations, the composite emission line profiles can display anywhere from 1 to 6 peaks.
 
Figure~\ref{fig:diversity} illustrates the diversity of shapes encountered in the profile database, calculated for different binary configurations. Each profile includes contribution from the primary and secondary mini-disks and the circumbinary disk. Individual profiles are broadened by rotational motion and random motion of the gas in the disk. Because the gas in the mini-disks has higher rotational velocity and is closer to the sources of continuum radiation, the emission line profiles contributed by the mini-disks often appear broader and stronger relative to the emission from the circumbinary disk. 

In Table~\ref{tab:peaks} we show the percentage of profiles characterized by a given number of peaks as a function of the orbital separation and eccentricity of the SBHB. One readily identifiable trend is that majority of profiles tend to have 1--3 peaks. The profiles with 5 and 6 peaks are relatively rare and entirely absent from SBHB systems with large orbital separations. This can be understood because SBHBs on tight orbits are characterized by compact mini-disks with high orbital velocities about the binary center of mass, both of which give rise to broad and multi-peaked lines in the wavelength space. Another trend is that SBHBs on eccentric orbits tend to have profiles with a higher number of peaks relative to the circular binaries with the same semi-major axis. This is because eccentric SBHBs sample a wider range of orbital velocities, allowing for a larger wavelength offset of individual components in the composite profile. 

As mentioned in Section~\ref{sec:cbd}, we assume that both accreting SBHs can shine as AGNs and illuminate all three disks in the system. In this setup, both mini-disks are illuminated by their central AGN as well as the off-center companion AGN. The illumination of the circumbinary disk by the two AGNs is always off-center. In Figure~\ref{fig:mutual} we show the effect of illumination of the primary mini-disk by the two AGNs (similar effect is present for the secondary mini-disk). The sequence of profiles at the top and bottom are created for two different SBHB configurations, arbitrarily chosen for illustration. 

The illumination by the secondary AGN resembles an off-center hot spot on the accretion disk surface. Figure~\ref{fig:mutual} shows that contribution to the total flux from such a hotpot sensitively depends on the alignment of the primary mini-disk with the binary orbit. Namely, when the two are coplanar ($\theta_1 = 0^{\circ}$), illumination due to the secondary AGN is negligible because of the small incidence angle of its photons on the mini-disk (first panel of Figure~\ref{fig:mutual}).  When the mini-disk and the binary orbit are misaligned even by a small amount, the illumination by the secondary AGN can make a significant contribution to the line flux (second panel of Figure~\ref{fig:mutual}). In the case of close binaries with highly misaligned mini-disks we find that this effect can increase the line flux up to several times (last panel). Depending on whether the hotspot moves away or toward the observer (as indicated by the azimuthal angle $\phi$) this extra flux may appear in the blue or the red wing of the emission line giving rise to an asymmetric profile (third and fourth panels). Therefore, the effect of illumination by a dual AGN can in principle be an indicator of the orbital alignment of the triple disk system, if it can be identified in the observed emission line profiles of candidate SBHB systems.

One more characteristic feature of the emission line profiles contributed by the triple disk system in our model is that the shape of a profile can change significantly over one orbital period of the binary. The centroids of the emission line profiles contributed by the mini-disks oscillate about the rest wavelength due to the orbital motion of the SBHs in the way similar to the spectroscopic stellar binaries. The emission from the circumbinary disk, which is anchored to the binary center of mass, is on the other hand centered on the rest wavelength of the system. As a result, a combination of the SBHB orbital motion and rotation of gas within each disk can produce complex and distinct features in SBHB systems relative to emission lines from stellar binaries.

\begin{figure*}[t]
\centering
\includegraphics[width=0.85\textwidth, clip=true]{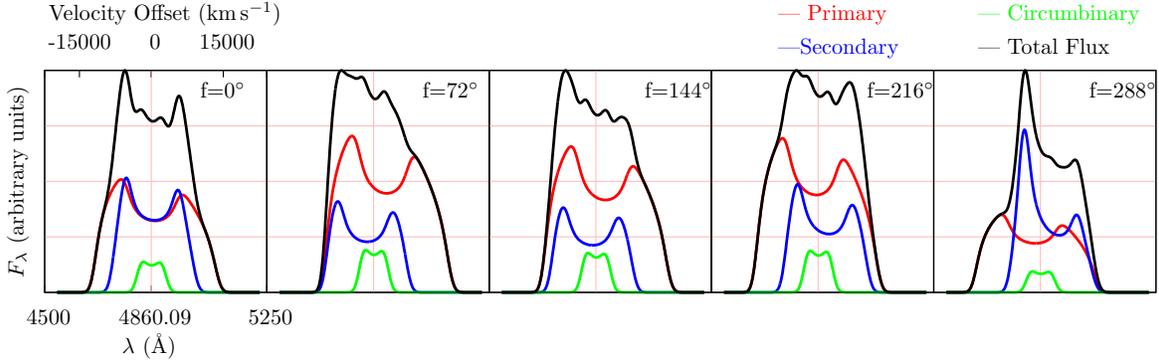}
\caption{Sequence of profiles showing temporal evolution of a composite emission line profile associated with an SBHB system described by  $q=1$, $a=5000\,M$, $e=0$, $R_{\rm in1}=500\,M_1$, $R_{in2}=1000\,M_2$, $i=75^{\circ}$, $\phi=0^{\circ}$, $\theta_1=0^{\circ}$, $\theta_2=15^{\circ}$, $\phi_1=0^{\circ}$, $\phi_2=300^{\circ}$. Total flux shown in arbitrary units (black line) is a sum of components contributed by the primary (red), secondary (blue) and circumbinary disk (green). SBHB orbital phase $f$ is marked in each panel.}
\label{fig:evolution}
\end{figure*}

Figure~\ref{fig:evolution} shows time evolution of a profile associated with an SBHB system in which profiles from both mini-disks are double peaked and asymmetric due to relativistic Doppler boosting (i.e., exhibit a higher blue shoulder). The same effect is also noticeable in the composite profile for all orbital phases except $f= 216^{\circ}$, when the blue and the red shoulder of the profile become comparable. At  $f=216^{\circ}$ the red wings of the two mini-disk profiles line up in wavelength giving rise to a relatively strong red peak in the composite profile. 
% The evolution happens on the SBHB orbital time scale and is a distinctive signature that is not expected to arise in the single black hole scenario. 

\subsection{Statistical properties of emission line profiles} \label{sec:stats}

The unique features of modeled emission line profiles associated with SBHB systems point to an intriguing possibility that, if it is possible to identify them in the observed SBHB candidates, these markers can be used to learn about the properties of the SBHBs. We analyze the trends in the modeled population of profiles by characterizing their shapes in terms of several commonly used distribution functions. These include the  location of the centroid (C), asymmetry index (AI), kurtosis index (KI), full width at half and quarter maximum (FWHM and FWQM), peak shift (PS), and centroid shift (CS). We use the following definitions:

\begin{eqnarray}
F &=&\sum_{i} F_{\lambda, i}\\
C &=&\frac{1}{F}\sum_{i} \lambda_i F_{\lambda, i}\\
\sigma^2 &=&\frac{1}{F}\sum_{i}(\lambda_i-C)^2 F_{\lambda, i}\\
\rm{AI} &=&\frac{1}{F\sigma^3}\sum_{i}(\lambda_i-C)^3 F_{\lambda, i} \label{eq_ai}\\
\rm{AIP} &=&(C-\lambda_m)/\sigma\label{eq_aip}\\
\rm{KI} &=&\frac{1}{F\sigma^4}\sum_{i}(\lambda_i-C)^4 F_{\lambda, i}\label{eq_ci}\\
\rm{FWHM} &=& \left[\lambda_r(1/2)-\lambda_b(1/2)\right]\frac{c}{\lambda_0}\label{eq_fwhm}\\
\rm{FWQM} &=& [\lambda_r(1/4)-\lambda_b(1/4)]\frac{c}{\lambda_0}\label{eq_fwqm}\\
\rm{PS} &=& (\lambda_p-\lambda_0)\frac{c}{\lambda_0}\label{eq_ps}\\
\rm{CS} &=&(C-\lambda_0)\frac{c}{\lambda_0}\label{eq_cs}
\end{eqnarray}

\noindent where $F_{\lambda, i}$ is the profile flux density at wavelength $\lambda_i$. The profile flux is normalized by the maximum flux measured at the peak wavelength, $\lambda_{p}$, so that $F_{\lambda}(\lambda_p)\equiv \max(F_{\lambda, i})=1$. $\lambda_b(x)$ and $\lambda_r(x)$ indicate the wavelength in the blue wing or the red wing of the profile, respectively, where the normalized flux drops to some level, $x$. $\lambda_0$ is the rest wavelength of the emission line and $\lambda_m$ is the median wavelength that divides profile into a half, so that 50\% of the flux lies to the left and to the right of it. The location of the profile centroid, $C$, is calculated as the flux weighted mean wavelength. 

We use two measures to characterize the asymmetry of the profiles: the asymmetry index (AI) and the Pearson skewness coefficient (AIP). The positive values of AI and AIP indicate profiles skewed toward short wavelengths (i.e., blue-leaning profiles) and the negative values indicate red-leaning profiles. However, AI and AIP calculated for the same profile sometimes have opposite signs, as they provide different measures of the profile asymmetry. Specifically, AI sensitively depends on the low intensity features in the profile wings, while AIP diagnoses the asymmetry in the bulk of the profile. 

We use the kurtosis index (KI), calculated as the fourth moment of the flux distribution, to evaluate the ``boxiness" of the profiles. By definition, the values of KI are always positive. Smaller values correspond to boxier profiles and larger values indicate cuspy profiles, with the top narrower than bottom. In addition, the relevant line widths, peak and centroid shifts are measured in units of velocity, as defined in equations~\ref{eq_fwhm} -- \ref{eq_cs}.

In calculation of all these statistical properties we adopt a cutoff at $F_c=0.01$ to mimic some fiducial level of spectral noise (but do not introduce actual fluctuations due to noise to the profiles). With ``noise" subtracted from the profile, we rescale the flux above the cutoff so that the maximum flux measured at the peak wavelength has the value of 1.0. We investigate the dependance of the distribution functions, characterizing the modeled profile shapes, on the value of $F_c$ in the Appendix~\ref{sec:appendixcutoff}.

\begin{figure*}[t]
\centering
\includegraphics[width=.85\textwidth, clip=true]{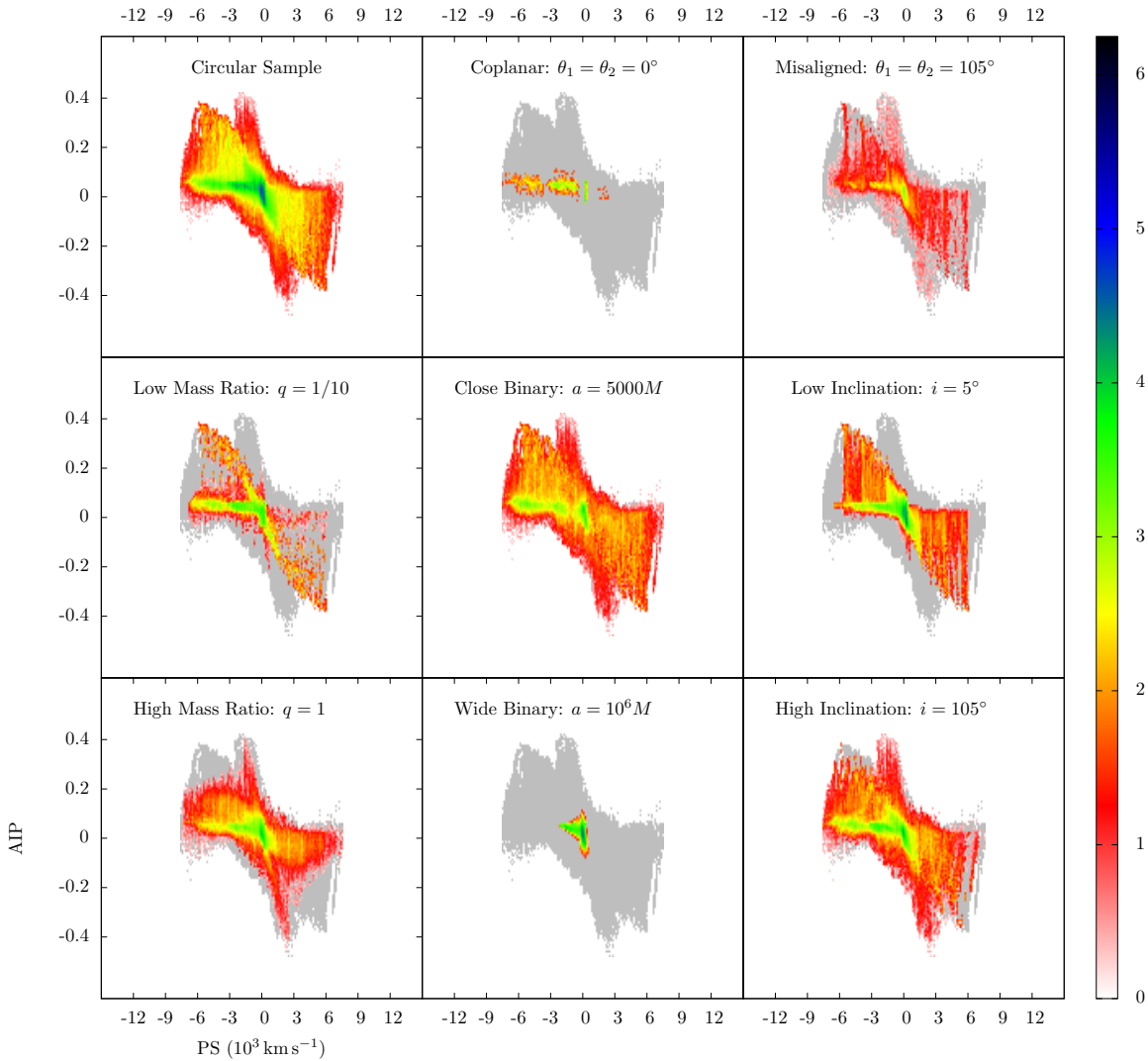}
\caption{AIP-PS map for profiles associated with circular SBHB systems (top left). Remaining panels display the distribution of profiles as a function of the alignment of the triple disk system ($\theta_1 = \theta_2 = 0^\circ$ and $105^\circ$), SBHB mass ratio ($q=1/10$ and 1), orbital separation ($a=5000\,M$ and $10^6\,M$), and inclination of the observer relative to the binary orbit ($i=5^\circ$ and $105^\circ$).  Color bar indicates the density of profiles (i.e., the number of profiles in each area element) plotted on $\log$ scale. Grey color outlines the footprint of the entire distribution shown in top left.}
\label{fig:phys1c}
\end{figure*}

\begin{figure*}[t]
\centering
\includegraphics[width=.85\textwidth, clip=true]{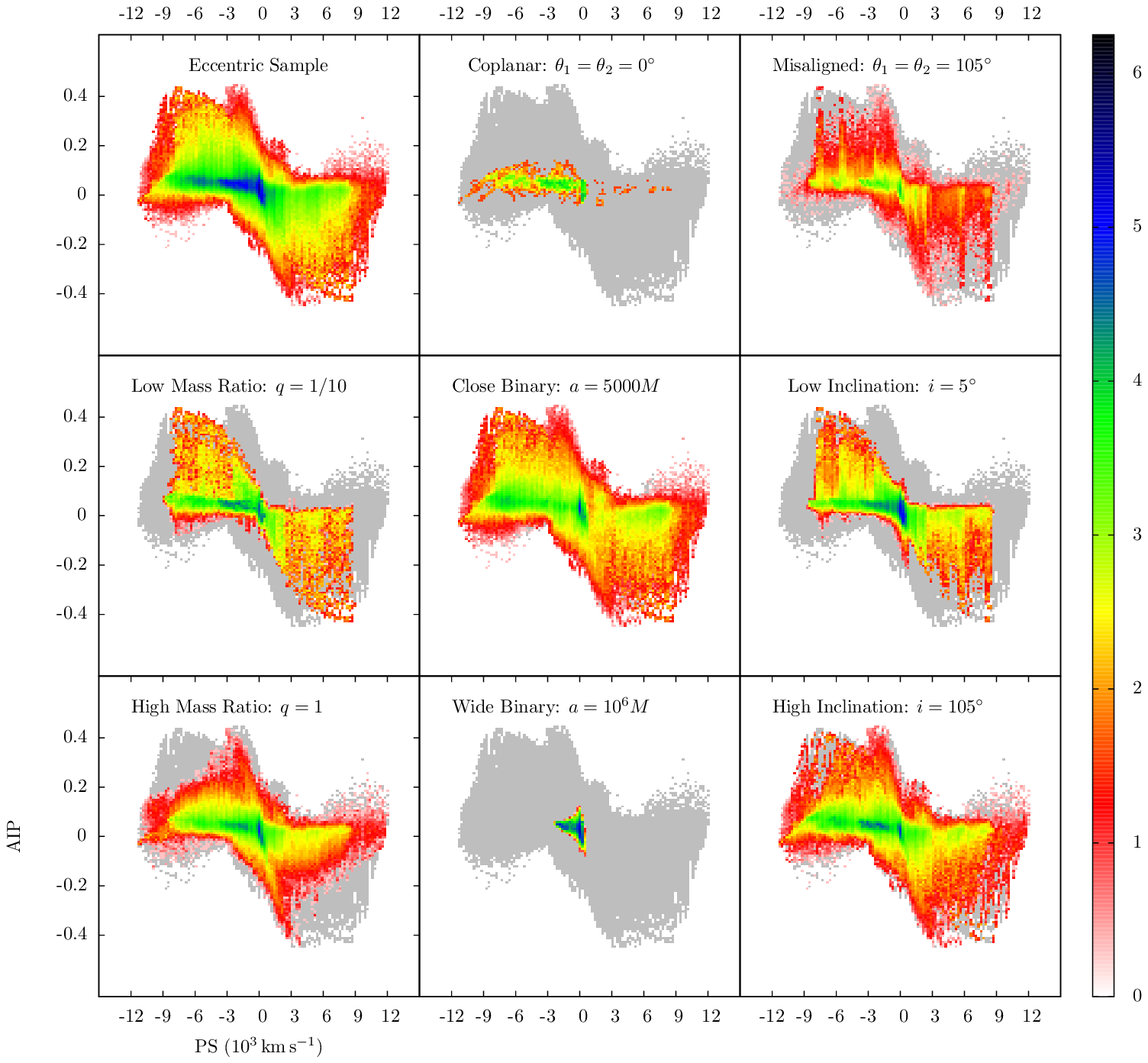}
\caption{AIP-PS maps for profiles associated with eccentric SBHB systems. Map legend is the same as in Figure~\ref{fig:phys1c}.}
\label{fig:phys1e}
\end{figure*}

\begin{figure*}[t]
\centering
\includegraphics[width=.75\textwidth, clip=true]{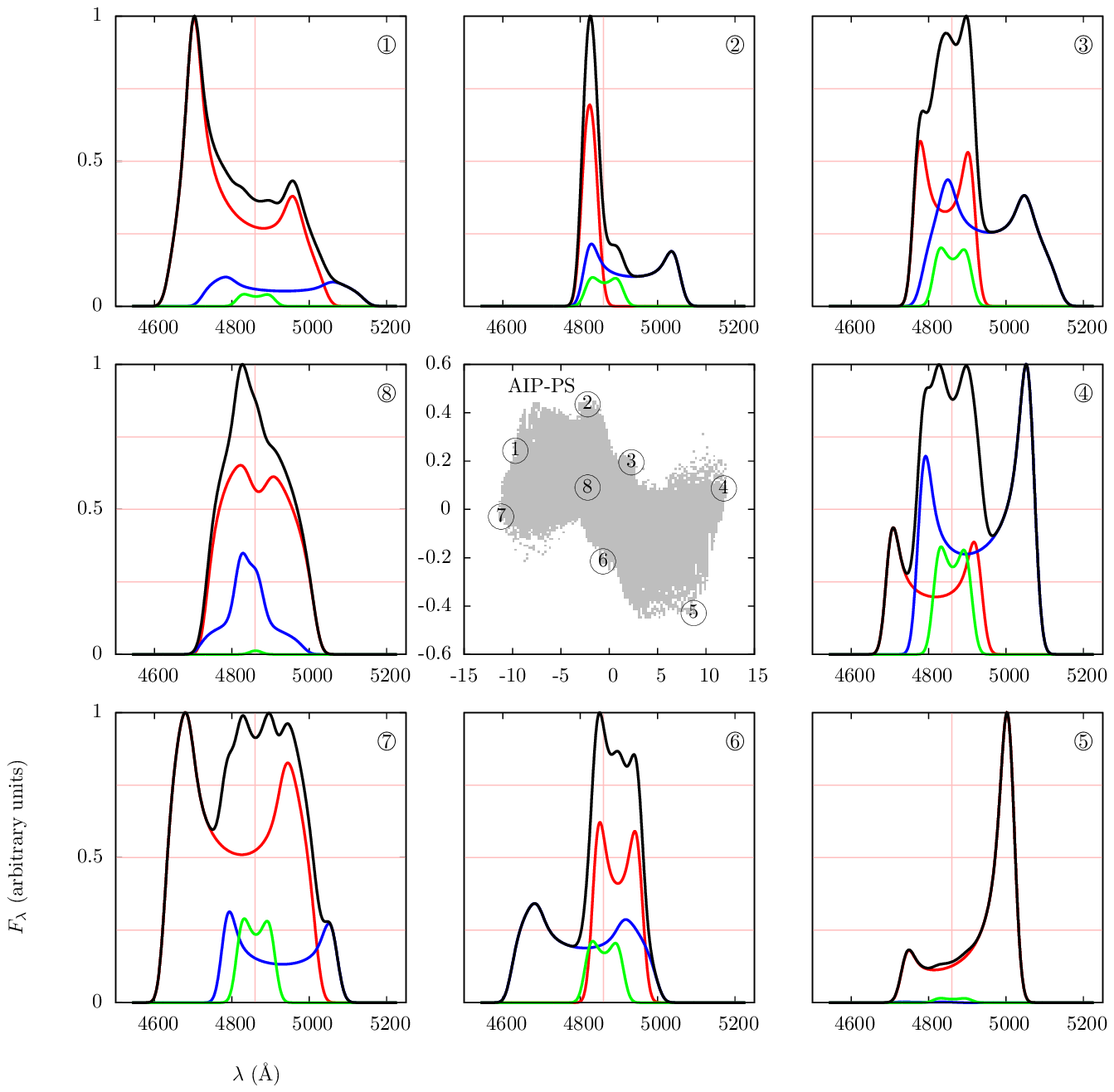}
\caption{Characteristic profile shapes occupying different regions in the AIP-PS parameter space. Central panel shows footprint of the AIP-PS distribution from the top left panel of Figure~\ref{fig:phys1e} with identical scale and labeling of axes. Pink vertical line in the outer panels marks the rest wavelength of the H$\beta$ emission line.}
\label{fig:stat1e}
\end{figure*}

We use statistical properties defined in equations~\ref{eq_ai} -- \ref{eq_cs} to construct a multi-dimensional parameter space of the emission line profiles and investigate their distribution as a function of the underlying SBHB parameters. In the remainder of the paper we visualize the multivariate distribution of profiles with 2-dimensional maps, which represent different projections through this parameter space. For example, in Figures~\ref{fig:phys1c} and \ref{fig:phys1e} we plot maps of AIP and PS values for profiles calculated for circular and eccentric binary configurations, respectively. The color marks the number density of profiles and indicates which portions of the parameter space are favored by the modeled profiles.

The top left map in Figures~\ref{fig:phys1c} and \ref{fig:phys1e} illustrates that AIP-PS distributions appear similar in the overall shape, with the eccentric sample having a wider range of the peak velocity shifts. This difference can be attributed to a wider range of orbital velocities sampled by eccentric binaries with the same semi-major axes. This topological similarity in the distribution of profiles from circular and eccentric SBHBs is present throughout the parameter space. Given the overlap, we plot only the distribution maps for the eccentric SBHBs in the rest of the paper and discuss any differences between the circular and eccentric samples in the text. 

Inspection of the top left panels in Figures~\ref{fig:phys1c} and \ref{fig:phys1e} reveals that a significant fraction of profiles are fairly symmetric (${\rm AIP} \approx 0$) and likely to exhibit the maximum peak at wavelengths shorter than the rest wavelength (${\rm PS}<0\,{\rm km\,s^{-1}}$). The latter is a consequence of the relativistic Doppler boosting, which for each individual disk preferentially boosts the blue shoulder of its emission line profile, creating an effect which is also noticeable in the composite profile. Another feature worth noting is that in both the circular and eccentric samples, the profiles that exhibit the strongest peak at shorter wavelengths are also preferentially blue-leaning and vice versa. In the next section, we inspect the remainder of the profile parameter space for similar trends and consider their relationship with the physical properties of the SBHB.

In Figure~\ref{fig:stat1e} we show examples of line profiles from different parts of the parameter space of Figure~\ref{fig:phys1e}, marked in the footprint of the map in the central panel. The shapes include profiles that exhibit symmetry (7 and 8), strong asymmetry (2 and 5), and large velocity offsets of the strongest peak (4 and 7).  As discussed in Section~\ref{sec:features}, the offset of the dominant peak towards longer wavelengths (evident in profiles 4 and 5) can occur in our model only under a specific circumstance: as a consequence of the illumination of a mini-disk by the companion AGN, when the hotspot is moving away from the observer. Inspection of profile 4 shows that the mini-disk with a strong hot spot is that around the secondary SBH (traced by the blue line) and around the primary SBH in profile 5 (traced by the red line). Moreover, profile 1 exemplifies the scenario where secondary illumination by the companion AGN dramatically boosts the blue wing of the profile from the primary mini-disk in configuration where the hotspot is moving towards the observer.

\subsection{Dependence of profiles on the physical parameters of the binary}\label{sec:dependence}

In this Section we investigate how the properties of modeled profiles vary as a function of the SBHB parameters, such as the alignment of the triple disk system, binary mass ratio, orbital separation, and inclination of the binary relative to the observer. We illustrate this dependence in the remainder of the panels in Figures~\ref{fig:phys1c} and \ref{fig:phys1e} where we show subsets of profiles associated with the specific value of SBHB parameter. These show that profiles from SBHBs with wide orbital separations ($a=10^6\,M$) tend to be very symmetric and concentrate in the center of the AIP-PS parameter space, while close binaries ($a=5000\,M$) have a much wider footprint. By implication, this means that only profile 8 shown in Figure~\ref{fig:stat1e} can be produced by systems with large orbital separations. 

Similarly,  any SBHB configurations where the mini-disks are co-planar with the binary orbit (and circumbinary disk, by assumption) are characterized by symmetric profiles with AIP$\approx 0$ with dominant peak shifted towards the blue part of the spectrum. The misaligned systems on the other hand are equally likely to be blue-leaning as well as red-leaning and reside in the range $-0.4\lesssim{\rm AIP} \lesssim 0.4$.  Therefore, profiles 1, 2, 3, 4, 5 and 6 cannot correspond to SBHBs with coplanar disks. As discussed in the previous paragraph, profiles 1, 4 and 5 also show strong contribution due to illumination by the companion AGNs, which is indeed expected to be most pronounced for configurations with misaligned disks. More generally, we find that the effect of illumination by the companion AGN is the main reason for difference between the AIP-PS distribution of profiles from coplanar and misaligned SBHB systems shown in Figure~\ref{fig:phys1e}.

On the other hand profiles associated with SBHB systems with different mass ratios ($q=1/10$ and 1) and different orientations of the binary orbit relative to the observer's line of sight ($\theta=5^\circ$ and $105^\circ$) show significant overlap in their distributions. Together, these plots indicate that the most important SBHB parameters that determine the degree of asymmetry and the position of the dominant peak in the emission line profile are the intrinsic alignment of the triple disk system and the orbital semi-major axis.

\begin{figure*}[t]
\centering
\includegraphics[width=.85\textwidth, clip=true]{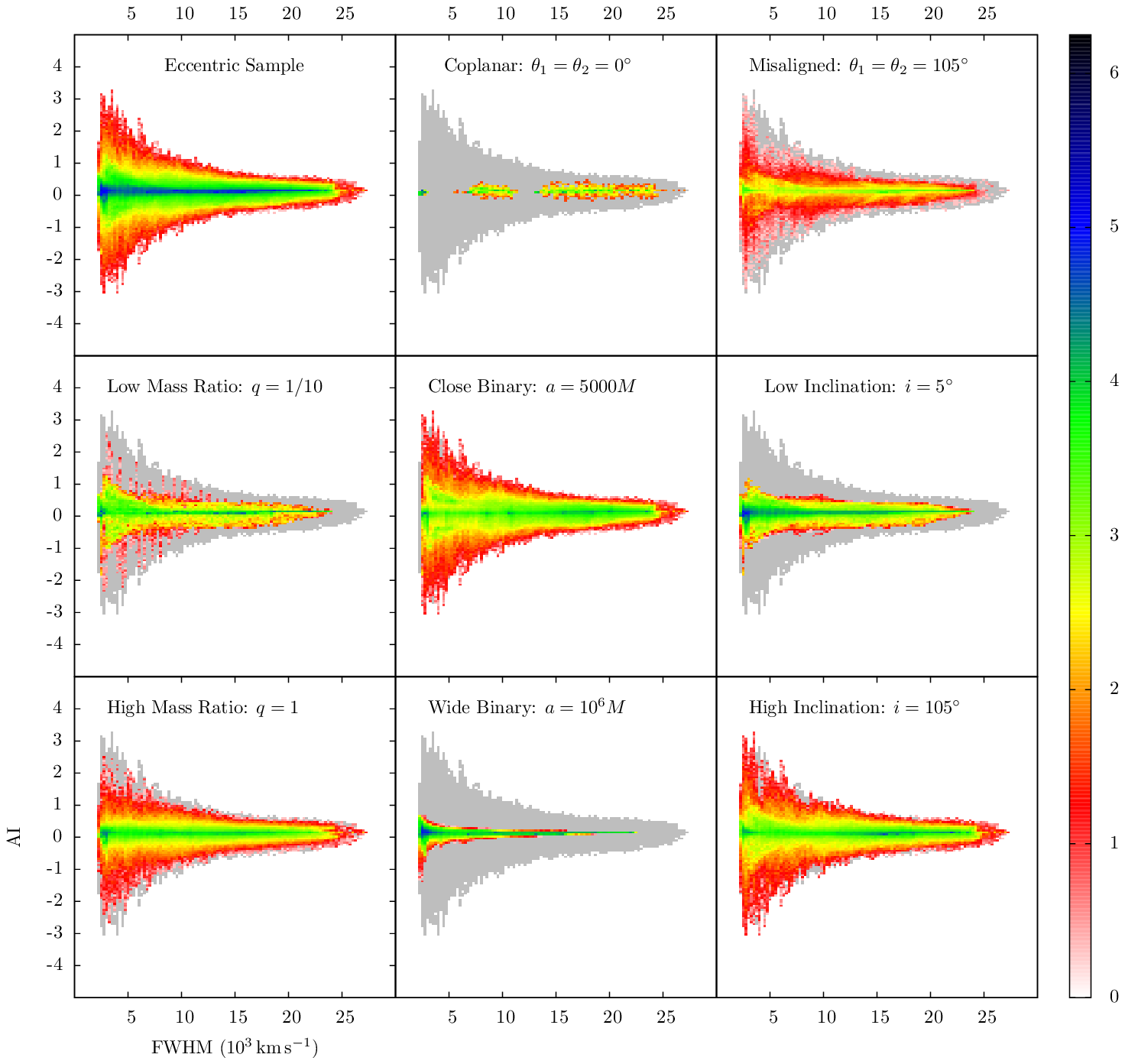}
\caption{AI-FWHM maps for emission line profiles associated with eccentric SBHB systems. Map legend is the same as in Figure~\ref{fig:phys1c}.}
\label{fig:phys2e}
\end{figure*}

\begin{figure*}[t]
\centering
\includegraphics[width=.75\textwidth, clip=true]{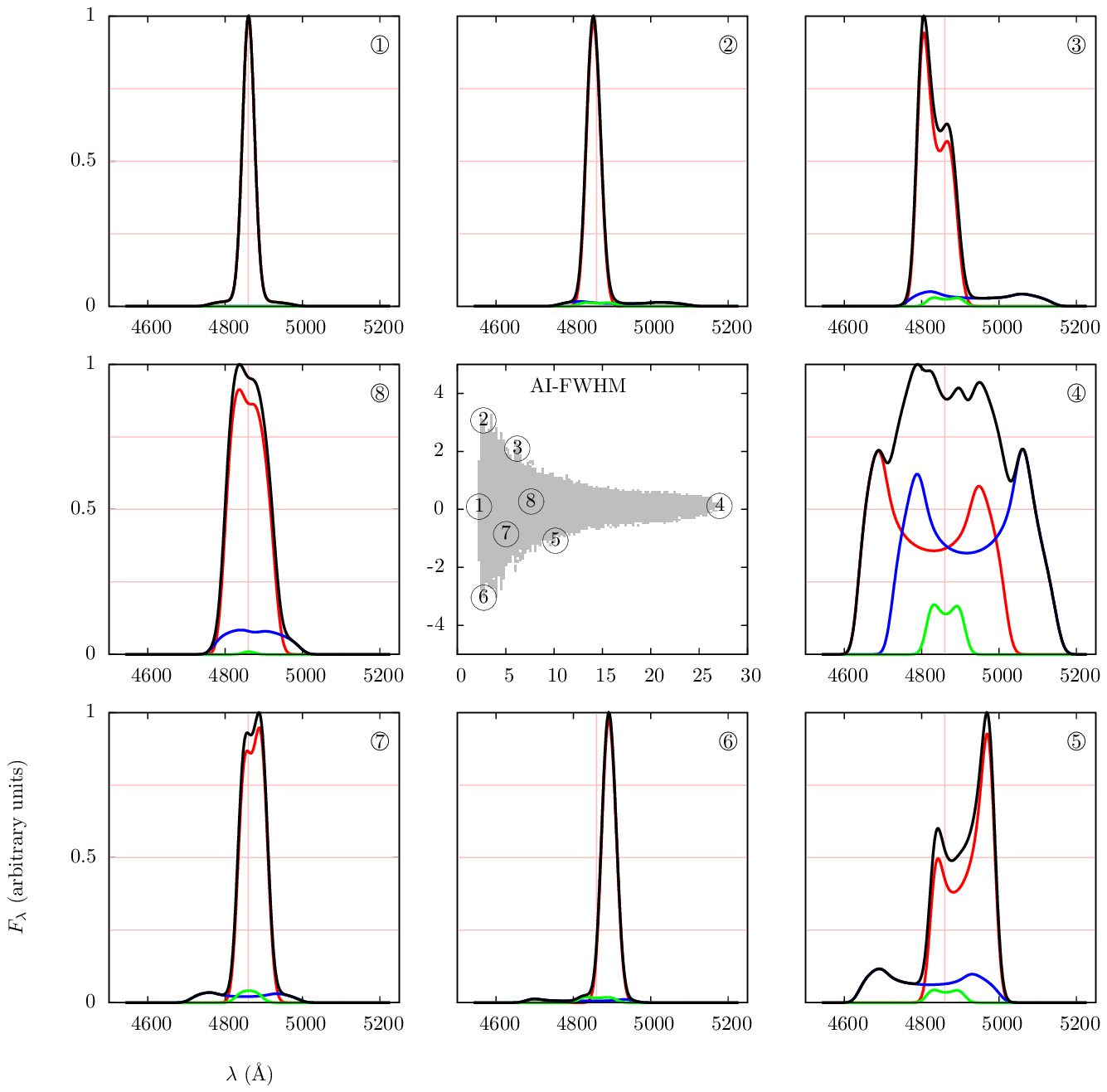}
\caption{Characteristic profile shapes occupying different regions in the AI-FWHM parameter space. Central panel shows footprint of the distribution from the top left panel of Figure~\ref{fig:phys2e} with identical scale and labeling of axes.}
\label{fig:stat2e}
\end{figure*}

\begin{figure*}[t]
\centering
\includegraphics[width=.85\textwidth, clip=true]{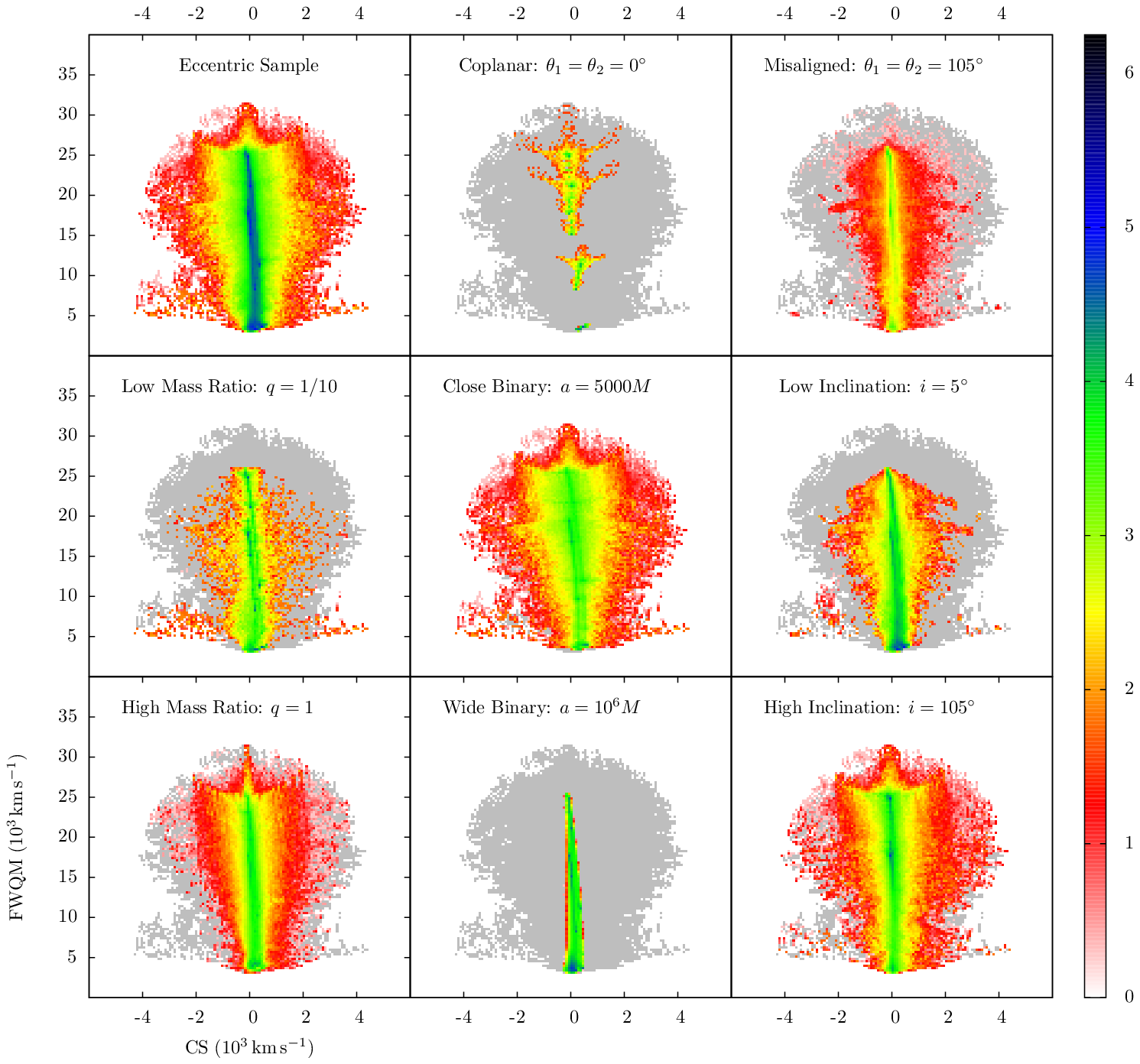}
\caption{FWQM-CS maps for profiles associated with eccentric SBHB systems. Map legend is the same as in Figure~\ref{fig:phys1c}.}
\label{fig:phys3e}
\end{figure*}

\begin{figure*}[t]
\centering
\includegraphics[width=.75\textwidth, clip=true]{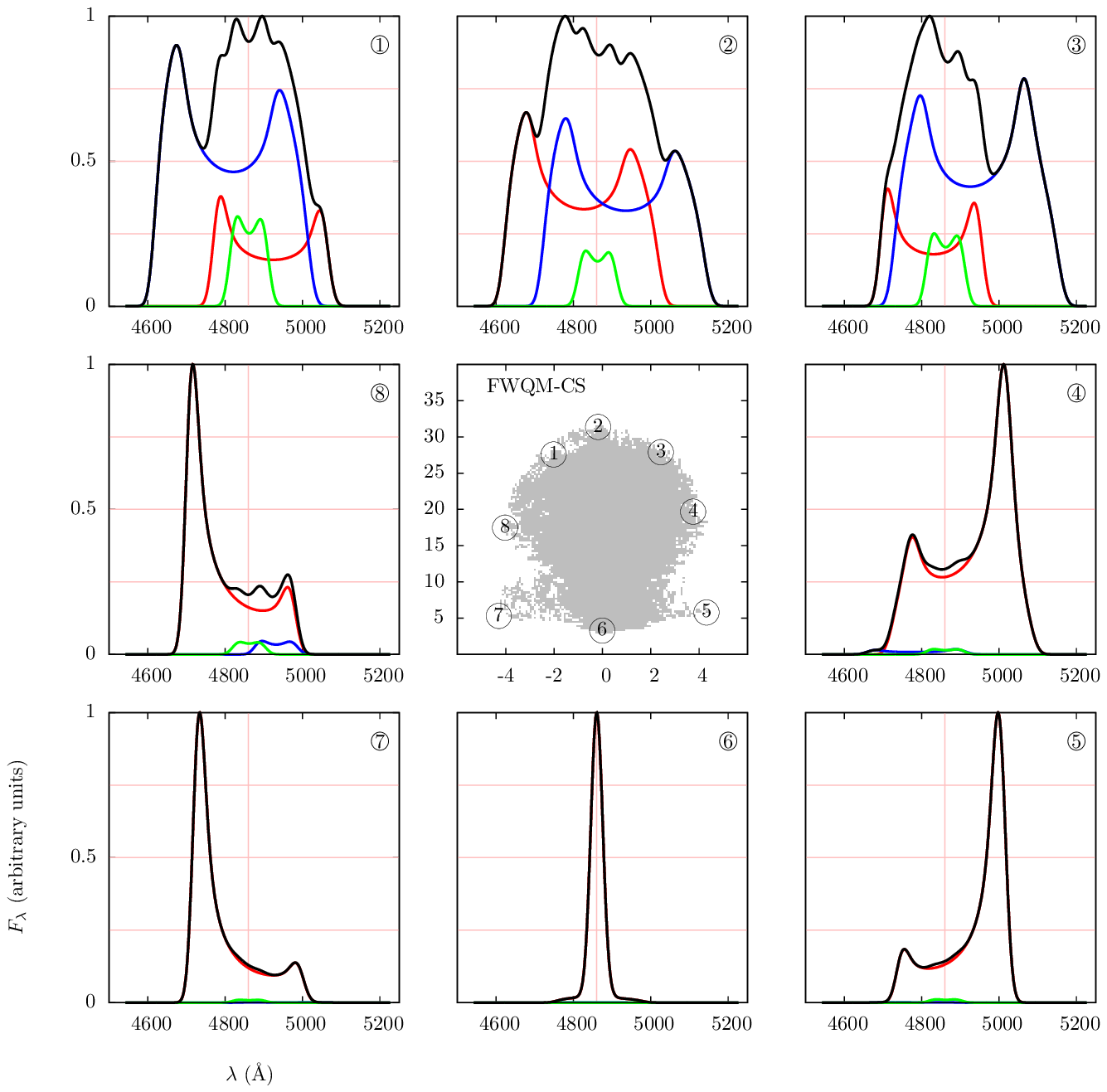}
\caption{Characteristic profile shapes in the FWQM-CS parameter space. Central panel shows footprint of the distribution from the top left panel of Figure~\ref{fig:phys3e} with identical scale and labeling of axes.}
\label{fig:stat3e}
\end{figure*}

\begin{figure*}[t]
\centering
\includegraphics[width=.85\textwidth, clip=true]{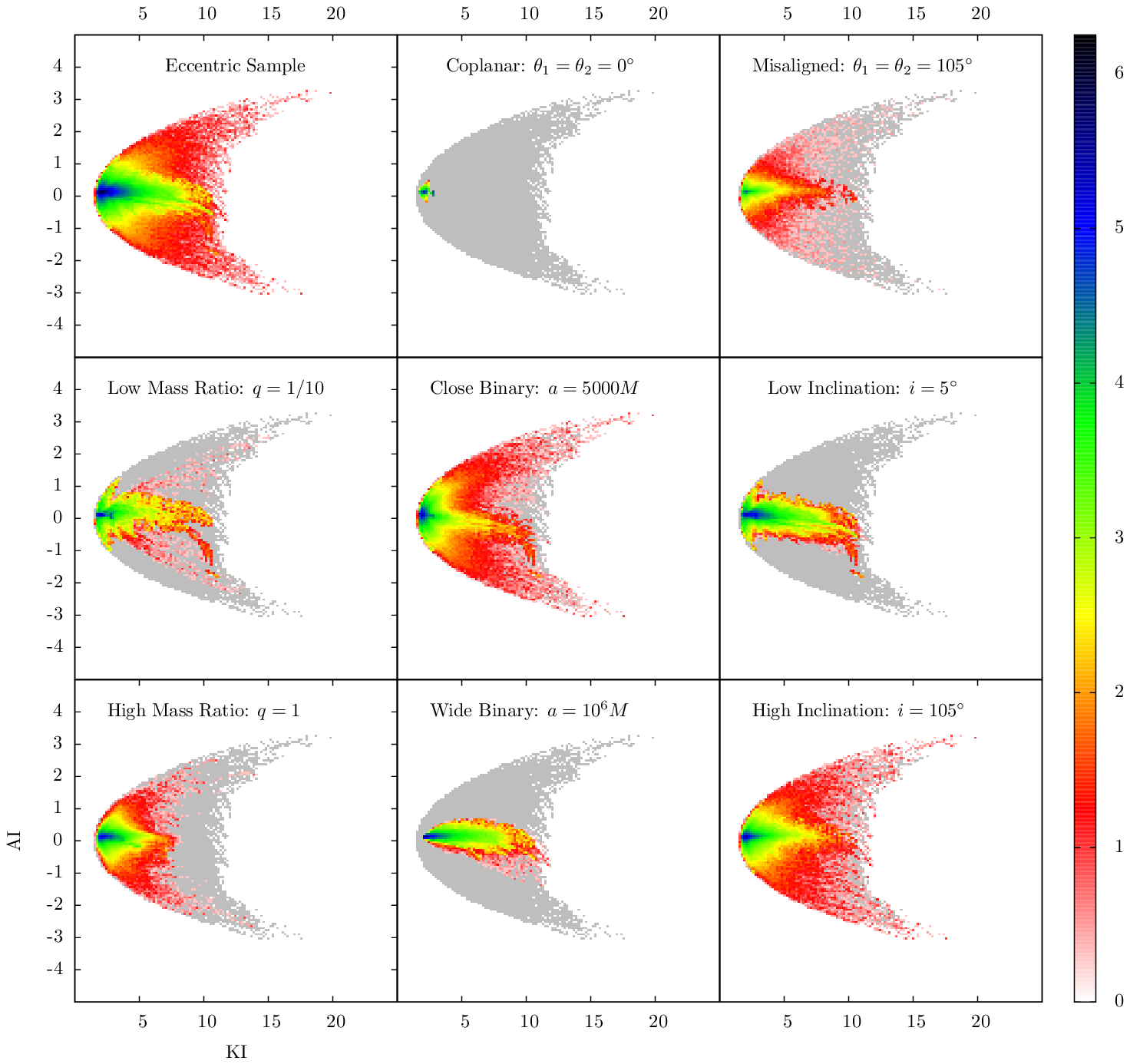}
\caption{AI-KI maps for profiles associated with eccentric SBHB systems. Map legend is the same as in Figure~\ref{fig:phys1c}.}
\label{fig:phys4e}
\end{figure*}

\begin{figure*}[t]
\centering
\includegraphics[width=.75\textwidth, clip=true]{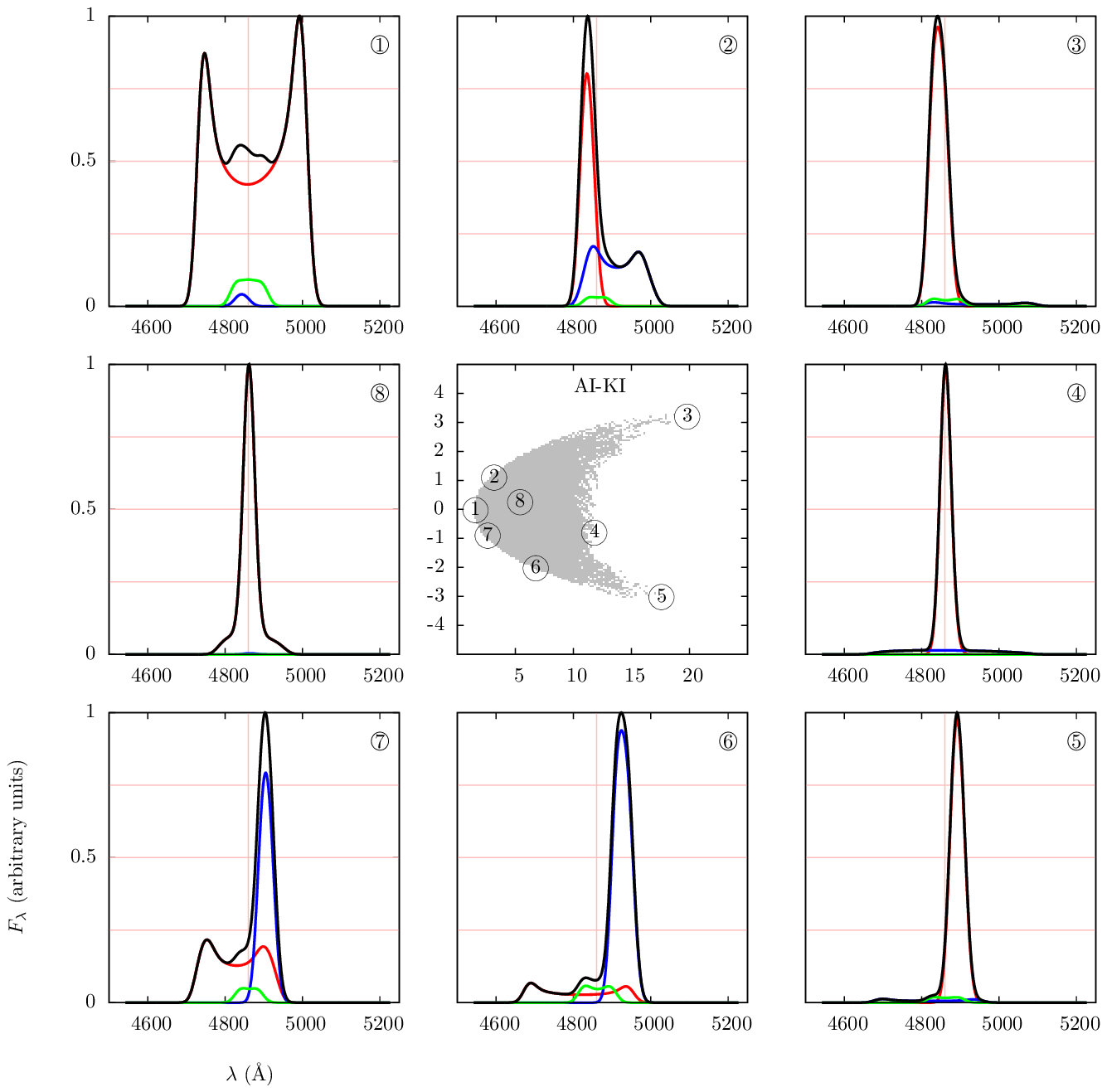}
\caption{Characteristic profile shapes in the AI-KI parameter space. Central panel shows footprint of the distribution from the top left panel of Figure~\ref{fig:phys4e} with identical scale and labeling of axes.}
\label{fig:stat4e}
\end{figure*}

\begin{figure*}[t]
\centering
\includegraphics[width=.8\textwidth, clip=true]{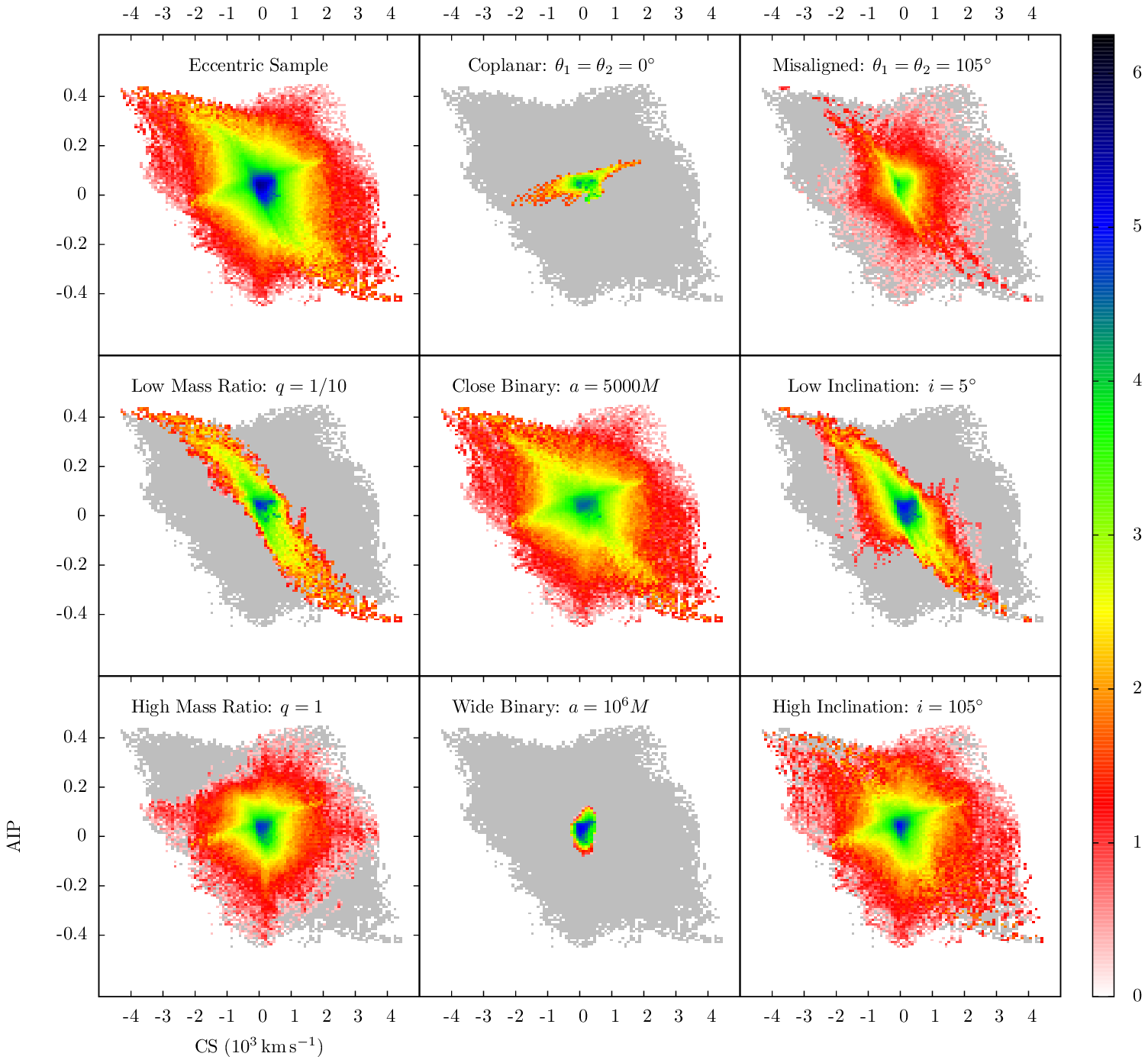}
\caption{AIP-CS maps for profiles associated with eccentric SBHB systems. Map legend is the same as in Figure~\ref{fig:phys1c}.}
\label{fig:phys5e}
\end{figure*}

\begin{figure*}[t]
\centering
\includegraphics[width=.8\textwidth, clip=true]{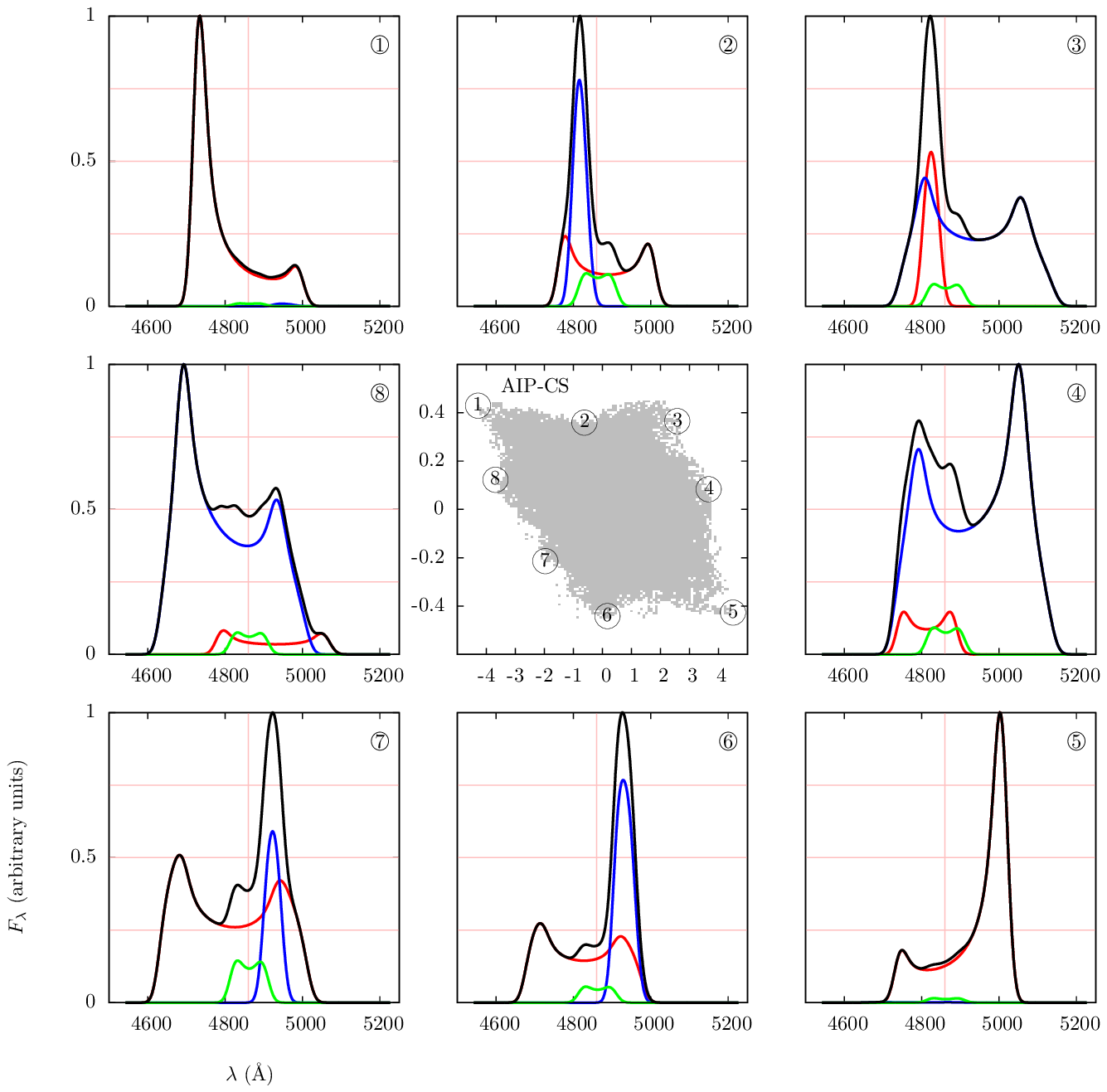}
\caption{Characteristic profile shapes in the AIP-CS parameter space. Central panel shows footprint of the distribution from the top left panel of Figure~\ref{fig:phys5e} with identical scale and labeling of axes.}
\label{fig:stat5e}
\end{figure*}

In the remaining figures we show four more projections of the multi-dimensional parameter space of the modeled emission line profiles including AI-FWHM (Figures~\ref{fig:phys2e} and \ref{fig:stat2e}), FWQM-CS (Figures~\ref{fig:phys3e} and \ref{fig:stat3e}), AI-KI (Figures~\ref{fig:phys4e} and \ref{fig:stat4e}) and AIP-CS maps (Figures~\ref{fig:phys5e} and \ref{fig:stat5e}). While more maps (i.e., parameter combinations) can in principle be constructed for this parameter space we focus on those that show distinct statistical distributions for any given SBHB property.

Figure~\ref{fig:phys2e} shows AI-FWHM maps for the eccentric sample of binaries, where we used formulation of the asymmetry index defined in equation~\ref{eq_ai}. The figure illustrates that profiles in the synthetic database have a wide range of FWHM values that extend to $28,000\,{\rm km\,s^{-1}}$ for eccentric sample of SBHBs. In comparison, the circular sample of binaries (not shown) is characterized by somewhat narrower profiles and ${\rm FWHM} < 23,000\,{\rm km\,s^{-1}}$. This difference can again be attributed to a wider range of orbital velocities sampled by eccentric binaries.

Similar to Figure~\ref{fig:phys1c} this map shows that SBHB systems with coplanar disks and SBHBs on wide orbits tend to produce symmetric profiles with ${\rm AI} \approx 0$, distinct from misaligned systems and close binaries. Furthermore, Figure~\ref{fig:phys2e} shows that low mass ratio systems ($q=1/10$) and those in which SBHB orbit is close to face-on orientation relative to the observer ($\theta=5^\circ$) also occupy a narrow range of $-1\gtrsim {\rm AI} \gtrsim 1$, relative to the footprint of the entire distribution. This means that a combination of AIP-PS and AI-FWHM maps can in principle be used to break the degeneracy between SBHBs with aligned disks or large $a$ and SBHBs with low inclination or low values of $q$.

Figure~\ref{fig:stat2e} shows the characteristic profile shapes occupying the AI-FWHM parameter space. Panels 2 and 6 illustrate the ability of AI to diagnose asymmetry in the low intensity features in profile wings even when the bulk of the profile is symmetric. By the same token, profile 5 has a lower value of AI than profile 6. This makes AI an useful diagnostic whenever the spectral noise level can be accurately determined and low intensity features clearly isolated. The profiles in panels 4 and 8 of Figure~\ref{fig:stat2e} have ${\rm AI} \approx 0$ showing that AI does not diagnose the asymmetry in the bulk of the profile. This tendency is the opposite from the AIP index, which makes them complementary diagnostics. Considering this in the context of the distributions discussed in the previous paragraph indicates that the low mass ratio SBHB systems and those in which SBHB orbit is close to face-on orientation relative to the observer can produce emission lines which are asymmetric in the bulk of the profile but show no significant asymmetry in the low intensity wings.

Figure~\ref{fig:phys3e} shows FWQM-CS maps calculated for emission line profiles from eccentric SBHB systems. Modeled profiles exhibit $|{\rm CS}| < 4,000\,{\rm km\,s^{-1}}$  and can have a broad base with ${\rm FWQM} < 30,000\,{\rm km\,s^{-1}}$. This figure illustrates that the location of the centroid is a strong function of $a$ in the sense that profiles from close binaries ($a=5000\,M$) can have a significantly wider range of CS values relative to the wide binaries ($a=10^6\,M$). Similarly the width of the profiles is most strongly affected by $q$ and $i$, where for low mass ratio binaries and nearly face-on systems ${\rm FWQM} < 25,000\,{\rm km\,s^{-1}}$,  lower than the entire sample of profiles. 

Figure~\ref{fig:stat3e} illustrates the diversity of profile shapes encountered in FWQM-CS parameter space. Interpreted together with Figure~\ref{fig:phys3e} it shows that the profiles 1, 2 and 3 must be produced by SBHB systems which satisfy either of these conditions: $q>1/10$, $\theta>5^\circ$ or $a\ll10^6\,M$. Similarly, profiles 1, 3, 4, 5, 7 and 8 cannot be associated with SBHB systems in which all disks are coplanar nor with wide binaries. Furthermore, profiles 5 and 7 cannot be associated with SBHB systems with $q=1$ and represent configurations where flux contributed by the primary mini-disk dominates over all other components. 
%More generally, these statements can be expressed in terms of likelihood that given profile corresponds to a given combination of SBHB parameters, where because of the overlap (degeneracy) in distributions some parameters can be statistically better constrained than others.

Figures~\ref{fig:phys4e} and \ref{fig:stat4e} show AI-KI maps and examples of the emission line profiles associated with eccentric SBHBs, respectively. A large fraction of profiles clusters around low values of AI and KI indicating a large number of symmetric and boxy shapes (see profiles 1 and 7). The AI-KI maps illustrate a strong dependence of the profile shapes on the alignment of the triple disk system, where aligned systems give rise to very boxy profiles with symmetric wings. Similarly, the asymmetry of the low intensity features in the profile wings (profiles 3 and 5) is a sensitive function of $a$ and $i$ but is less sensitive to $q$, because distributions for different values of the SBHB mass ratio overlap to a significant degree. 

The pair of Figures~\ref{fig:phys5e} and \ref{fig:stat5e} show the AIP-CS projection of the parameter space and the profile shapes in it, respectively. Similar to previous maps, the statistical distributions are a strong function of $a$, followed by the degree of triple disk alignment marked by the angles $\theta_1$ and $\theta_2$. This implies that AIP-CS maps can be used as a relatively sensitive diagnostic for these properties. The statistical distributions as a function of $q$ and $i$ are also distinct, so AIP-CS may also be used to constrain these parameters, although with a somewhat larger degree of degeneracy.

The profile shapes shown in Figure~\ref{fig:stat5e} are drawn from the rim of the AIP-CS distribution and are representative of shapes associated with close SBHBs, those with $q\approx 1$ (with the exception of 1 and 5) and large inclination. Profiles in panels 1, 4, 5 and 8 exhibit asymmetry due to one strongly dominant peak produced by illumination of one mini-disk by the companion AGN. As discussed in Section~\ref{sec:features}, this effect produces strong contribution to the flux of the composite profile in close binaries where the illuminated mini-disk is not aligned with the SBHB orbit, consistent with the binary properties shown in the AIP-CS maps. Profiles 2, 3, 6 and 7 also have a very pronounced, dominant peak which in their case arises due to incidental alignment of constituent profiles, rather than illumination by the other AGN.

In this section we analyzed the dependance of the modeled profile shapes on the key parameters describing SBHB and triple disk configurations. For convenience, we summarize the most important results below.
\begin{itemize}
\item The shapes of modeled emission line profiles are a sensitive function of the binary orbital separation. Compared to systems with small orbital semi-major axis, line profiles of wide SBHBs are more symmetric and occupy a relatively narrow range of values in terms of boxiness, peak and centroid shifts, and FWHM.
\item Similarly, modeled profile shapes are a sensitive function of the degree of alignment in the triple disk system. Line profiles associated with SBHB systems with nearly coplanar disks tend to be symmetric, boxy, and weakly affected by the secondary illumination from the companion AGN relative to the misaligned systems.
\item The synthetic emission line profiles tend to be less sensitive (or more degenerate with respect) to the SBHB mass ratio according to a majority of statistical distributions calculated in this work. The exception is the AIP-CS parameter space in which the low and high $q$ systems trace distinct correlations.
\item Besides the SBHB mass ratio, the modeled profile shapes show a comparable degree of degeneracy with respect to the binary orbital inclination relative to the observer. Compared to systems with high orbital inclination, line profiles of low inclination systems tend to be more symmetric, especially in the extended profile wings, and have somewhat lower values of FWHM. Similar to the SBHB mass ratio, the low and high inclination systems trace distinct correlations in the AIP-CS parameter space. 
\end{itemize}

%====================================================================================================%

\section{Discussion}\label{sec:discussion}

\subsection{Do modeled broad emission line profiles carry an imprint of SBHB parameters?}

The ultimate goal of this investigation is to investigate whether the broad emission line profiles, commonly used in spectroscopic observational searchers to select SBHB candidates, can be used to decode the properties of bona fide SBHBs. In this work we make a step in this direction by first examining whether modeled broad emission line profile shapes convey any information about the parameters of SBHBs and their BLRs. If so, further development of this and similar models is of interest, as well as a comparison of such models with the data.

The answer to this question is not obvious a priori: while composite profiles are potentially rich in information, the properties of SBHBs may be difficult to extract because of the complex emission geometry of multiple accretion disks. In practice, this means that any model designed to represent such systems must depend on a number of parameters and so do calculated emission line profiles (listed in Table~\ref{table:parameters} and Section~\ref{sec:params} for model presented here). Because of the dependence of profiles on multiple parameters and their degeneracy, it is unlikely that a unique match between a model and an observed SBHB can be achieved by attempting to fit the observed profile with arbitrary parameter combinations. 

This argues for an approach based on statistical distributions as a more promising way to analyze profile shapes. In this approach observed profiles can be matched to the modeled database based on their values of AIP, KI, FWHM, PS and CS. Each observed profile would map into a subset of modeled profiles with similar statistical properties that represent different modeled SBHB configurations. This correspondence of one observed profile and multiple SBHB configurations is a direct manifestation of degeneracy of the SBHB parameters. As a result, one could make a statement about the likelihood that the observed profile corresponds to some given SBHB configuration. If instead of one, a temporal sequence of observed profiles from the same SBHB is available for comparison with the modeled database, this could further help to reduce degeneracy. 

Our results indicate that the modeled profiles show distinct statistical properties as a function of the semi-major axis, mass ratio, eccentricity of the binary, and the degree of alignment of the triple disk system. In our model the SBHB systems on eccentric orbits are more likely to produce broader emission line profiles and complex profiles with multiple peaks relative to the circular cases. Thus, an observed profile compared to the synthetic database can be assigned a finite probability in the context of this model that it originates with the circular or eccentric SBHB based on its shape (see however the discussion below). 
 
Furthermore, mini-disks in smaller separation binaries which are misaligned with the binary orbital plane are subject to strong illumination by both AGNs in the system. As a consequence of the off-center illumination, such systems give rise to very asymmetric profiles that can exhibit significant peak or centroid velocity shifts. This is the dominant reason why all statistical distributions shown in this work are sensitive functions of parameters that control orbital separation and disk alignment. Indeed, in our model these two features of SBHB systems are most easily discernible based on profile shapes.

In comparison, the effects of the binary mass ratio and SBHB orientation relative to a distant observer on profiles shapes are most discernible in the AIP-CS distribution (for both $q$ and $i$) and AI-KI distribution (for $i$ and to a lesser degree $q$). It is interesting to note that SBHBs with low $q$ or nearly face-on orbits ($i\approx 0^\circ$) tend to show a significant degree of correlation between the Pearson's skewness coefficient and the peak or centroid velocity shifts (Figures~\ref{fig:phys1e} and \ref{fig:phys5e}), where the AIP-CS correlation is more pronounced. This implies that such SBHB systems give rise to specific asymmetric profiles. As the offset of the dominant peak increases, the profile becomes more asymmetric resulting in red leaning profiles with the strongest peak shifted towards red or the blue leaning profiles with the strongest peak shifted towards blue. 

 Visual inspection of such profiles indicates that their shapes tend to be strongly affected by the off-center illumination of the primary mini-disk, which dominates the flux in the composite profile. This can be understood because in our model the emission from the primary mini-disk typically dominates over that from the secondary and circumbinary disks for the smallest values of $q$. Even so, the composite emission line profile does not default to a fairly symmetric double peaked profile from an accretion disk about a single SBH, precisely because of the illumination by the secondary AGN.

It follows that the most characteristic features of the modeled profiles presented here are a direct consequence of the presence of multiple BLRs (giving rise to profiles with multiple peaks) and their illumination by two AGN, both of which are an inherent property of the SBHB model. The distinct statistical distributions suggest that SBHB properties are indeed imprinted in the population of modeled profiles, albeit with some degeneracy, which for any given SBHB parameter can be statistically quantified. Based on this we conclude that models of broad emission line profiles from SBHBs in circumbinary disks can have predictive power and as such merit further investigation. 

\subsection{Can modeled emission line profiles be compared with the observed profiles from SBHB candidates?}

The next relevant question is whether the modeled emission line profiles presented here can be directly compared with those from spectroscopically selected SBHB candidates. We make several such comparisons below but note that they do not constitute a proof that the observed candidates are indeed SBH binaries.

Examination of the observed emission line profiles from SBHB candidates presented by \citet{eracleous12}, \citet{decarli13}, \citet{liu13} and \citet{li16}, shows that these profiles can be asymmetric and offset but are usually quite smooth and characterized by one or two peaks, unlike some of the profiles in our database with complex structure and up to 6 peaks. Admittedly, the fraction of profiles with such high number of peaks is relatively small in our database and they are more common for eccentric binaries. On the other hand, the modeled single and double-peaked profiles are most common in SBHB configurations with semi-major axes $a\geq 5\times 10^4\,M$, which for a $10^8\,M_\odot$ binary translates to $\geq 0.25$\,pc. Therefore, if comparison between the observed and modeled profiles is made at the face value, it would favor moderately wide bound binaries.

Note that both AIP-PS and AIP-CS correlations have been reported in spectroscopically targeted SBHB candidates and have not been found in a  control sample of matching AGN \citep{eracleous12, runnoe15}. A qualitative comparison of the observed sample in Figure~18 of \citet{runnoe15} with our modeled AIP-PS distributions in Figures~\ref{fig:phys1c} and \ref{fig:phys1e} shows that they cover a similar range of AIP values but that the observed profiles tend to have peak velocity shifts in a narrow range between $-4000\lesssim {\rm PS} \lesssim 3000\,{\rm km\,s^{-1}}$. In the context of the SBHB model this disfavors configurations of SBHBs with either the smallest or the widest orbital separations and favors moderately wide binaries and binaries with misaligned disks.

Another correlation identified in the sample of observed SBHBs by the same authors is between the third moment of the flux distribution of profiles\footnote{Related to the properties calculated in this work as $\mu_3 = \sigma^3 {\rm AI}$.}, $\mu_3$, and FWHM. Namely, \citet{eracleous12} report that the values of $\mu_3$ in the observed emission line profiles tend to decrease with increasing FWHM. This correlation is not seemingly present in our modeled sample regardless of the adopted parameter cut. As noted before however, the value of AI (and that of the related parameter $\mu_3$) sensitively depends on the noise level, which in observed profiles is very likely different from the fiducial noise level we adopt in our calculations of AI. We will take this difference into account in future work, when we carry out a more detailed comparison between the observed and modeled samples.
 
 Along similar lines, the FWHM measured by \citet{eracleous12} in their sample of 88 candidates reaches up to $18,000\,{\rm km\,s^{-1}}$. Our synthetic profiles are however characterized by values of FWHM as high as $23,000\,{\rm km\,s^{-1}}$ for the circular and $28,000\,{\rm km\,s^{-1}}$ for the eccentric sample of binaries. The modeled profiles are therefore inherently wider than those observed, regardless of the SBHB parameter cut. 
 %A visual comparison of the observed and modeled AI-FWHM distributions leads to the following interpretation in the context of the SBHB model: the observed profiles could not have been predominantly emitted by the binary systems with aligned disks, those with the widest orbital separations, or face-on orientation of the binary orbit.

The tendency of modeled profiles to exhibit richer and more diverse structure can to some degree be ascribed to their dependence on the semi-major axis, as discussed at the beginning of this section, or perhaps a larger degree of ``smoothing" in real profiles due to either the presence of noise or a larger velocity dispersion of the emitting gas on average. On the other hand a mismatch in the range of measured FWHM between the two populations cannot be trivially explained. The FWHM measured in modeled profiles is a function of the orbital velocity of the gas in each disk and the orbital velocity of the binary, both of which are inherent characteristics of SBHB systems. If anything, increasing the velocity dispersion of the gas ($\sigma$) in our model, in order to produce smoother profiles, would result in even wider profiles and more tension between the observed and modeled samples.

A qualitative comparison therefore highlights some intriguing similarities and also points to differences between the two samples. The former motivate further development of models of broad emission line profiles from SBHB systems, given their potential to interpret profiles from observed bona fide SBHBs. The latter may arise either due to a true difference between the two samples of profiles or because physical processes giving rise to the broad optical emission line profiles were not entirely captured by our model. It is therefore important before attempting more detailed comparisons to examine the impact of any simplifying assumptions made in the current model.

\subsection{Simplifying assumptions and their implications}\label{secsec:approx}

Perhaps the most important physical mechanism that can significantly modify the appearance of the spectrum and emission lines is the radiative feedback from the binary AGN, capable of driving winds and outflows from the circumbinary region. Several recent simulations of SBHBs in circumbinary disks indicate that despite strong binary torques, accretion into the central cavity continues more-less unhindered relative to the single SBH case \citep{dorazio13,farris14,shi15}. This point is of particular interest because AGN feedback from an accreting binary SBH can considerably change the structure, thermodynamic and ionization properties of the circumbinary region. 
 
In this work, we assume that the emissivity of each broad line region arises due to photoionization by the two AGNs but neglect the effects of radiation pressure on the dynamics and optical depth of the emitting gas. In order to address this, in Paper II we will calculate the profiles of low-ionization emission lines by generalizing models that account for radiative transfer effects through a disk wind \citep{cm96, mc97, flohic12, chajet13}. These works have demonstrated that disk models that reproduce relatively rare double-peaked emitters can also describe emission from BLRs of most AGNs once these effects are accounted for, pointing to their broad applicability.

The key effect of the accretion disk wind is to modify the shape of a broad emission line profile. This occurs because the radiation pressure from the central AGN lifts-off the low density gas from the surface of the disk and launches it along streamlines above the disk. The photons (in this case H$\beta$) escaping from a single accretion disk encounter increased optical depth through the emission layer and as a consequence, the peaks of an initially double peaked profile move closer and eventually merge, producing a narrower single peaked profile. 

Comparisons of such single peaked disk-wind model profiles with emission lines from a set of SDSS quasars show that observed lines are consistent with moderately large optical depth in the disk wind and indicate that most AGNs may be subject to this type of feedback \citep{flohic12}. It is therefore reasonable to assume that if SBHs in a binary can accrete at rates comparable to the general population of AGNs, they are likely to produce similar effects. The reprocessing of the H$\beta$ photons through the accretion disk wind may indeed produce smoother and narrower profiles in better general agreement with the observed sample of SBHB candidates and AGNs in general. However, the same effect may also ``wash out" some of the characteristic features encountered in our modeled profiles, thus weakening their dependence on the properties of the binary. We will assess the diagnostic power from broad emission line profiles affected by the accretion disk wind and associated with SBHBs in circumbinary disks in Paper II.

In addition to the accretion disk wind the emissivity can also be modified by shocks, impacts of streams from the circumbinary disk onto the accretion disks around the individual SBHs \citep{roedig14b} and by the presence of overdense lumps that may form in the inner region of the circumbinary disk \citep{farris14}. These features have been predicted by some theoretical models and simulations and if indeed present in binary accretion flows, they would increase the complexity of the emission line profiles by creating hot spots and localized regions of high emissivity. The presence, persistance and exact emission properties of these features however sensitively depend on thermodynamics of the SBHB accretion flow, which remains to be understood and at the present cannot be derived from first principles. We do not account for contribution to the emissivity of the broad lines from shocks and overdensities but note that they can be added to the existing model should that be necessary.
 
Another approximation used in our model is that the two mini-disks, as well as the circumbinary disk are circular in shape. In this scenario, the outer edges of the mini-disks and the inner edge of the circumbinary disk are determined by SBHB tidal forces and are not free parameters of the model (see Section~\ref{sec:params}). Simulations however show that the mini-disks and the circumbinary disk can exhibit varying degrees of eccentricity as a consequence of tidal deformation by the binary, an effect which is most pronounced for comparable-mass binaries \citep{farris14}. From the stand point of our semi-analytic model this implies that additional parameters may be required in order to describe the geometry of the emission regions around SBHBs, leading to additional degrees of freedom in profile shapes. The distinct property of an elliptical accretion disk is that it can naturally give rise to double peaked emission line profiles in which the red peak is stronger than the blue, a feature that cannot be reproduced by a circular disk \citep{eracleous95}. In our model this type of asymmetry is present in less than 50\% of the profiles and it arises in two ways: either through summation of individual profiles which results in a stronger red peak (see for example panels 6 and 7 in Figure~\ref{fig:stat5e}) or due to illumination by the companion AGN (panels 4 and 5 in Figure~\ref{fig:stat5e}). We therefore reproduce such an effect even though we only consider circular BLRs in our model. If our model accounted for elliptical disks the database may contain a larger fraction of profiles with the dominant red peak (reflected in the positive value of the peak velocity shift) but at the expense of a number of additional parameters.

An additional assumption used in our model is that of prograde binaries. Namely, motivated by theoretical works described in Section~\ref{sec:intro} we assume that the SBHB and the circumbinary disk are coplanar and rotate in the same sense. At the same time, the mini-disks are allowed to assume arbitrary orientation (and sense of rotation) relative to the SBHB orbit. A circumbinary disk with an arbitrary orientation relative to the SBHB orbit would however still produce a single- or a double-peaked profile that is centered on the SBHB rest frame, similar to the profiles shown in this work. Since the total flux of the composite profile is dominated by the primary and secondary mini-disks, the assumption about co-planarity of the circumbinary disk should not strongly affect our results. Note however that simulations of retrograde SBHBs in circumbinary disks show a different dependence of SBH accretion rates on orbital eccentricity \citep{roedig14} from that assumed in Equation~\ref{eq:mscale} of this work. This is another ingredient that can in principle be added to the model, if counter-rotating SBHB configurations are of interest.

Because we evaluate Doppler boosting and gravitational redshift in the weak field limit and neglect bending of light  (see Appendix~\ref{sec:appendixsingle}) we can only faithfully calculate the emission line profiles that arise in configurations in which the photons are emitted far from the immediate environment of black holes (i.e., at distances larger than tens of Schwarzschild radii) and in which they do not travel on grazing trajectories over the SBHs. Both of these requirements are satisfied in our model given the assumed sizes of emission regions and the fact that we do not allow for edge-on configurations characterized by the disk inclination angles in the range $80 - 100^{\circ}$.  Along similar lines, we do not account for lensing of one mini-disk by the companion SBH when the two SBHs are in conjunction. Such configurations are expected to be rare and short lived and should not significantly affect the overall statistical distribution of the emission line profiles.

The parameter values in Table~\ref{table:parameters} are chosen so as to provide a relatively uniform but not necessarily dense coverage of the SBHB parameter space. This can be seen in the middle top panel of Figure~\ref{fig:phys3e}, where "branches" at ${\rm FWQM} \approx 22,000\,{\rm km\,s^{-1}}$ and  $27,000\,{\rm km\,s^{-1}}$ carry an imprint of the underlying SBHB parameter choices, most likely that of the binary orbital inclination. Because of the extent of the parameter space, the number of sampled configurations quickly adds up to nearly 15 million, even with a handful of choices per parameter. While this rate of sampling may be acceptable for surveying the properties of emission line profiles, a denser coverage can be obtained for sub-regions of the parameter space.

It is worth noting that other physical processes can potentially mimic the emission signatures of SBHBs discussed here. These include but are not limited to the recoiling SBHs \citep{blecha16} and local and global instabilities in single SBH accretion disks that can give rise to transient bright spots and spiral arms \citep{flohic08,lewis10}. In that sense, the model described in this paper can be used to interpret observed emission line profiles in the context of the SBHB model but cannot be used to prove that they originate with veritable SBHB systems. For example, profiles of SBHB candidates observed in multiple epochs can be compared against the synthetic database individually, in order to determine the likelihood distribution for underlying SBHB parameters for each profile independently. The entire time series of observed profiles can also be compared against the time series of matching modeled profiles as an added consistency check for the inferred SBHB parameters. We defer this type of analysis to future work.

%====================================================================================================%

\section{Conclusions and Future Prospects}\label{sec:conclusions}

This work is motivated by advances in observational searches for SBHBs made in the past few years which are represented by better designed, multi-wavelength and multi-year observational campaigns. Observational challenges notwithstanding, spectroscopic searches for SBHBs seem capable of delivering statistically significant sample of binary candidates and their first results are broadly consistent with theoretical predictions. While selection of a well defined sample of SBHBs remains a principal goal in this research field, an equally important and timely consideration is what can be learned once such sample is available. In this context we develop a model to calculate the broad emission line profiles from SBHBs in circumbinary disks guided by a wealth of theoretical results in the literature. In this work, which constitutes part I of the series, we consider whether complex, composite emission line profiles from SBHB systems can be used as a diagnostic of the binary properties. 

We use the SBHB model to calculate a database of 14.8 million emission line profiles arising from a triple disk system associated with the binary. In this setup each disk acts as a BLR and contributes emission line flux resulting in a broad composite profile. We analyze the  modeled emission line profiles in terms of the commonly used statistical distribution functions in order to determine their dependence on the underlying binary parameters. We find that the modeled profiles show distinct properties as a function of the binary semi-major axis, eccentricity, mass ratio, alignment of the triple disk system and orientation relative to the observer.  The most characteristic features of modeled profiles are a direct consequence of the presence of multiple BLRs and their illumination by two accretion powered SBHs, both of which are a unique property of the SBHB model. Thus, models of broad emission line profiles from SBHBs in circumbinary disks have predictive power and can in principle be used to infer distribution of these parameters in real binaries.
 
We identify some intriguing similarities between the observed SBHB candidates and our synthetic profiles. Both groups exhibit correlation between the Pearson skewness coefficient and the peak or centroid velocity shift. Initial comparison of the two samples at the face value favors SBHB candidates which are moderately wide binaries with misaligned disks. On the other hand, the database of modeled profiles contains more diverse profile morphologies and on average wider profiles than the observed sample of SBHB candidates or a general population of AGNs. This suggests that not all relevant physical phenomena are fully captured by our model, a question that given the potential of this and similar models merits further investigation. 

The leading contender for a physical mechanism that can modify the appearance of the emission lines is radiative feedback from the binary AGN, capable of driving winds and outflows in the circumbinary region. We will investigate the importance of this mechanism in future work, where we will calculate the emission line profiles by taking into account the radiative transfer effects through a disk wind. More specifically, we will re-evaluate the diagnostic power of broad emission lines and carry out a direct comparison of the observed and modeled profile samples.

We conclude by noting that the emission signatures discussed here may not be unique to SBHB systems and that there is a possibility that they can be mimicked by other physical processes, driven by local and global instabilities in single SBH accretion disks. The model described in this work is therefore a promising tool that can be used to interpret the observed emission line profiles in the context of the SBHB model but should not be considered a conclusive test of binarity.

%====================================================================================================%

\acknowledgements
We are grateful to Michael Eracleous and Jessie Runnoe for their insightful and useful comments and thank the anonymous referee for a thoughtful report. This research was supported in part by the National Science Foundation under Grant No. NSF AST-1211677 and by the National Aeronautics and Space Administration under Grant No. NNX15AK84G issued through the Astrophysics Theory Program. T.B. acknowledges support from the Research Corporation for Science Advancement through a Cottrell Scholar Award. T.B. is a member of the MAGNA project (\url{http://www.issibern.ch/teams/agnactivity/Home.html}) supported by the International Space Science Institute (ISSI) in Bern, Switzerland. Numerical simulations presented in this paper were performed using the high-performance computing cluster PACE, administered by the Office of Information and Technology at the Georgia Institute of Technology.

%====================================================================================================%

\appendix
\label{appendix}

\section{A: Broad Emission Line Profiles from Circular Keplerian Disk} \label{sec:appendixsingle}

\begin{figure*}[t]
\centering
\includegraphics[width=.5\textwidth, clip=true]{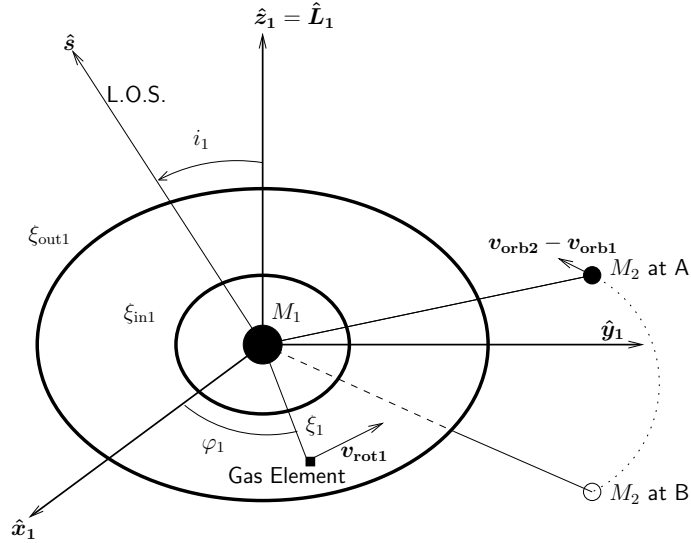}
\caption{Geometry of the system in which two sources can illuminate the disk around an SBH. In this illustration the coordinate system is centered on the primary SBH defined so that the $z_1$-axis is parallel to the angular momentum vector of the gas in the disk and the line of sight $\boldsymbol{\hat s}$ belongs to the $x_1z_1$ plane.  The illumination of the disk by the secondary AGN can be seen by a distant observer only when the secondary AGN belongs to the same half-plane with the observer (location A) and is otherwise blocked by the disk (location B).}
\label{fig:single}
\end{figure*}

We describe each disk in the triple disk system as a circular Keplerian, geometrically thin accretion disk in the weak-field approximation as outlined by \citet{chen89}, \citet{chen89b} and \citet{eracleous95}. Specifically, we use implementation that assumes optically thin emission from the skin of the disk \citep[equation~19 in][]{chen89} and neglect bending of light in gravitational field of an SBH \citep[encoded in equation~8 of both][]{chen89b,eracleous95}. We first outline the key elements of this model (hereafter referred to as the {\it single disk model}) and then describe modifications we made in order to calculate emission line profiles from triple disk systems. In the single disk model the flux of the broad emission line profile measured in the observer's frame can be expressed as an integral over the surface of the emitting disk defined in terms of the properties in the disk frame:
\begin{equation}
\label{eq:flux1}
F(\nu_{\rm obs})=\frac{M^2\nu_0}{d^2} 
\int\limits_{0}^{2\pi} \int\limits_{\xi_{\rm in}}^{\xi_{\rm out}}   
I(\xi,\nu_{\rm turb}) D_{\rm rot}^3 \left( 1-\frac{2}{\xi} \right) ^ {-\frac{1}{2}} \xi\,  d\xi\, d\varphi
\end{equation}
where $M$ is the mass of the central object, $\nu_0$ is the rest frequency of the emission line, $d$ is the distance from the center of the disk to the observer, $\xi=r/M$ is the radius in the disk in dimensionless units, $\xi_{\rm in}$ and $\xi_{\rm out}$ are the inner and outer edge of the emission region, respectively, and  $\varphi$ is the azimuthal angle measured in the plane of the disk (note that $\varphi$ is different from $\phi$ defined in Section~\ref{sec:params}). The geometry of such a disk is illustrated in Figure~\ref{fig:single}.

$I(\xi,\nu_{\rm turb})$ is the specific intensity of light emitted at radius $\xi$ and frequency $\nu_{\rm turb}$ 

\begin{equation}
\label{eq:intensity1}
I(\xi,\nu_{\rm turb})= \frac{1}{4\pi} 
\frac{\epsilon(\xi)}{(2\pi)^{1/2} \sigma} \, 
\exp[{-(\nu_{\rm turb}-\nu_0)^2/2\sigma^2}]
\end{equation}
where $\epsilon(\xi)$ is the disk emissivity as a function of radius. In the single disk model $\epsilon = \epsilon_{0} \cdot \xi^{-p}$ represents the emissivity of the disk illuminated by a single, central source.  The emissivity constant, $\epsilon_{0}$, is proportional to the luminosity of the photoionizing source, which we assume is powered by accretion onto an SBH. For the purposes of this calculation we therefore assume that $\epsilon_{0}\propto \dot{M}$. Geometric arguments, as well as photoionization calculations, indicate that $p \approx 3$ is a reasonable value for the emissivity index \citep{csd89} and we adopt it in our calculations. In the next section we extend this formalism to account for illumination of the disk by two AGNs, associated with two SBHs as illustrated in Figure~\ref{fig:single}.

The frequency of photons emerging locally from the disk ($\nu_{\rm turb}$) is shifted by turbulent motion of gas and assumed to have a Gaussian distribution about the rest frequency $\nu_0$ \citep{chen89b}.  We adopt a characteristic velocity dispersion of the gas due to turbulent motion of $\sigma \simeq 850 \, \rm km\,s^{-1}$, which corresponds to a characteristic frequency shift of $\Delta \nu/ \nu_0 = 850\, {\rm km\, s^{-1}} / c \simeq 2.8 \times 10^{-3}$, where $c$ is the speed of light.  Our choice of the characteristic velocity dispersion is motivated by the values inferred from radio-loud AGN with double-peaked emission lines, which have been successfully modeled with either the circular or elliptical disk models and turbulent broadening in the range of $\sim 600-3000 \, \rm km\,s^{-1}$ \citep{eh94, eh03, strateva03}. 

Several additional effects can impact the frequency of emitted photons, including the rotational motion of gas in the disk, relativistic Doppler boosting, and gravitational redshift. Classically, the Doppler factor associated with the motion of the gas in the disk can be expressed as
\begin{equation} \label{eq:dopfactor3}
D_{\rm cl}=\frac{1}{1- \boldsymbol{v}_{\rm rot} \boldsymbol{\cdot \hat{s}}} = (1+\xi^{-1/2}\sin i \sin \varphi)^{-1}
\end{equation}
where $\boldsymbol{v}_{\rm rot} = \xi^{-1/2}\, \boldsymbol{\hat{e}_{\phi}}$ is the velocity vector of a given surface element of the disk in units of the speed of light, $\boldsymbol{\hat{s}}$ is the unit vector along the line of sight of the observer, and $i$ is the inclination of the disk with respect to the observers line of sight, as illustrated in Figure~\ref{fig:single}. The two relativistic effects can be accounted for by adding terms for the special relativistic beaming, $D_{\rm sr}=\sqrt{1-v_{\rm rot}^2}$, and general relativistic gravitational redshift, $D_{\rm gr}=\sqrt{1-2/\xi}$, yielding
\begin{equation}
\label{eq:dopfactor1}
D_{\rm rot} = \left(1-\frac{1}{\xi}\right)^{1/2} \left(1-\frac{2}{\xi}\right)^{1/2} D_{\rm cl}
\simeq \left(1-\frac{3}{\xi} \right)^{1/2} D_{\rm cl}
\end{equation}
where we used the weak-field approximation (valid when $\xi \gg 1$) to obtain the final expression. For an emission element in the disk located at $\xi=500$ the shift in the frequency of emitted light due to the relativistic effects can be estimated as $D_{\rm rot} \simeq 0.99699 D_{\rm cl}$. For the $H\beta$ transition this amounts to nearly 15\AA, an offset that is in principle detectable given the spectral resolution of optical surveys and therefore should be accounted for in the model. The relativistic Doppler factor can then be expressed as 
\begin{equation} 
\label{eq:dopfactor2}
D_{\rm rot}=\nu_{\rm obs}/\nu_{\rm turb}=(1-3/\xi)^{1/2}(1+\xi^{-1/2}\sin i \sin \varphi)^{-1}.
\end{equation}
where $\nu_{\rm obs}$ marks the frequency of the photon measured by the observer. Our derivation of equation~\ref{eq:flux1} departs from that of \citet{chen89b} and \citet{eracleous95} because it does not include relativistic bending of light and is thus applicable under two conditions. The first is that the photons are emitted by a gas element far away from the black hole ($\xi \gg 1$). The second is that emitted photons do not travel on ``grazing" orbits over the black hole. Our calculation satisfies both by having the $H\beta$ photons emerge from the radii in the disk $\xi \geq 500$ and by eliminating edge-on configurations characterized by inclination angles of the disk in the range $80 - 100^{\circ}$.

%------------------------------------------------------------------------------------------------------------
\section{B: Broad Emission Line Profiles From a Triple Disk System}\label{appendixbinary}

In this section we describe modifications to the single disk model introduced in order to calculate the emission line profiles from the triple disk system associated with an SBHB. This is accomplished in three steps in which we: (a) define the orientation of the three disks relative to the orbital plane of the SBHB and relative to a distant observer (b) evaluate the emissivity of each disk illuminated by the two AGNs and (c) sum the three components of flux to calculate the composite emission line profile in the frame of the binary. 

\subsection{Geometry of the triple disk system}

In order to determine the mutual orientation of the three disks and the binary orbit, as well as their relative orientation with respect to the observer's line of sight, we define three coordinate systems, each anchored to the center of its resident disk as in Figure~\ref{fig:single}. The coordinate system associated with the circumbinary disk coincides with that associated with the binary orbit and has the origin in the SBHB center of mass.  We refer to it as the {\it binary} or {\it SBHB frame} in the rest of the text. In order to distinguish among the properties calculated in these reference frames we introduce subscripts where ``1" and ``2" correspond to the primary and secondary mini-disks, and ``3" to the circumbinary disk, respectively. Furthermore, because we carry out the calculation of flux in dimensionless, geometric units (as shown in the previous section), the subscripts also indicate that distances are measured in units of $M_1$ and $M_2$ in the frames of the primary and secondary mini-disks, and $M=M_1+M_2$ in the frame of the binary.

Figure~\ref{fig:binary} illustrates the coordinate system anchored to the binary orbital plane where the SBHB center of mass marks the origin and the $z$-axis points in the direction of the orbital angular momentum of the binary, directed out of the page. The $x$-axis points towards the pericenter of the primary SBH orbit and is parallel to the orbital semi-major axis of the binary, $a=(a_1+a_2)$.  We describe the orientation of the two SBHs in the orbital plane of the binary as a vector pointing from the primary to the secondary SBH (see Figure~\ref{fig:binary})
\begin{equation}
\label{eq:ell}
\boldsymbol{l}= l\; \boldsymbol{\hat{e}}_r = l(\cos{f} \, \boldsymbol{\hat{x}}+ \sin f \: \boldsymbol{\hat{y}})
\end{equation}
where $l=a(1-e^2)/(1 - e \cos f)$ is the separation of the AGNs, $e$ is the orbital eccentricity, $f$ is the orbital phase of the SBHB measured counter-clockwise from the $x$-axis to the instantaneous location of the secondary SBH, and $\boldsymbol{\hat{e}}_r$ is the unit vector parallel to $\boldsymbol{l}$. 

We define the orientation of the observer in the SBHB frame with a vector $\boldsymbol{\hat{s}}=\sin i \cos\phi  \,\boldsymbol{\hat{x}} + \sin i \sin\phi \,\boldsymbol{\hat{y}} + \cos i \,\boldsymbol{\hat{z}}$. The inclination angle, $i$, describes the orientation of the observer's line of sight relative to the vector of orbital angular momentum of the SBHB. For example, the inclination angle $i=0^\circ$ represents the clockwise binary seen face-on and values $i>90^\circ$ represent counter-clockwise binaries. The azimuthal angle $\phi$ is measured in the binary orbital plane, from the positive $x$-axis to the projection of the observer's line of sight, in counter-clockwise direction. For circular SBHBs varying the orbital phase $f$ is equivalent to varying the azimuthal orientation of the observer and in this case we adopt a single nominal value of $\phi=0^\circ$ in calculation of the emission line profiles. This is however not the case for the eccentric binaries, in which case $f$ and $\phi$ take independent values.

We define the orientation of the primary mini-disk by specifying the orientation of its rotation axis (given by the unit vector of the disk angular momentum, $\boldsymbol{\hat{L}_1}$) in terms of the polar and azimuthal angles $\theta_1$ and $\phi_1$ measured in the SBHB frame: 
\begin{equation}
\boldsymbol{\hat{L}_1}= \sin\theta_1 \cos\phi_1 \;\boldsymbol{\hat{x}} + \sin\theta_1 \sin\phi_1 \;\boldsymbol{\hat{y}} + \cos\theta_1 \;\boldsymbol{\hat{z}} \;\; .
\end{equation}
Equivalently, we use $\theta_2$, $\phi_2$ to specify the orientation of the secondary disk, given by the unit vector of the disk angular momentum, $\boldsymbol{\hat{L}_2}$. As described in Section~\ref{sec:params} mini-disks are coplanar with the SBHB orbit when $\theta_1=\theta_2=0^{\circ}$ and the gas in the mini-disks exhibits retrograde motion relative to the SBHB when $\theta_i>90^\circ$. The azimuthal angles $\phi_1$ and $\phi_2$ are measured in the binary orbital plane, from the positive $x$-axis to the projections of the rotation axes of the mini-disks, in counter-clockwise direction. 

With known orientations of the mini-disks and the observer in the SBHB frame we can evaluate the inclinations of the mini-disks with respect to the observer's line of sight
\begin{equation}
\label{eq:i1}
\cos i_1= \boldsymbol{\hat{L}_1} \cdot \boldsymbol{\hat{s}} = \sin i \cos\phi \sin\theta_1 \cos\phi_1 + \sin i \sin\phi \sin\theta_1 \sin\phi_1 + \cos i \cos\theta_1
\end{equation}
The orientation of the secondary AGN relative to the primary mini-disk can be calculated as follows
\begin{eqnarray}
&&\sin \theta_{M2} \cos\phi_{M2}=
\boldsymbol{\hat{e}_r}\cdot \boldsymbol{\hat{x}_1}=
\boldsymbol{\hat{e}_r} \cdot 
\left( \frac{\boldsymbol{\hat{L}_1} \times \boldsymbol{\hat{s}}}{\sin i_1} \times \boldsymbol{\hat{L}_1} \right)\\
&&\sin \theta_{M2} \sin\phi_{M2}=
\boldsymbol{\hat{e}_r} \cdot  \boldsymbol{\hat{y}_1}=
\boldsymbol{\hat{e}_r} \cdot \frac{\boldsymbol{\hat{L}_1} \times \boldsymbol{\hat{s}}}{\sin i_1}\\
&&\cos \theta_{M2}=
\boldsymbol{\hat{e}}_r \cdot \boldsymbol{\hat{L}_1} =
\sin \theta_1 \cos \phi_1 \cos f + \sin \theta_1 \sin \phi_1 \sin f=
\sin\theta_1 \cos (\phi_1 - f)
\end{eqnarray}
where $\theta_{M2}$ and $\phi_{M2}$ are the spherical polar coordinates describing the location of the secondary AGN in the primary mini-disk frame. Equivalent expressions can be written for the secondary mini-disk by replacing subscript ``1" with ``2"
\begin{eqnarray}
&&\cos i_2= \boldsymbol{\hat{L}_2} \cdot \boldsymbol{\hat{s}}\\
&&\sin \theta_{M1} \sin\phi_{M1}=- \boldsymbol{\hat{e}_r } \cdot \boldsymbol{\hat{y}_2}\\
&&\cos \theta_{M1}= - \boldsymbol{\hat{e}_r} \cdot \boldsymbol{\hat{L}_2}
\end{eqnarray}
where the minus signs in the last two equations encode the opposition of the primary and secondary SBHs relative to the center of mass. In all configurations we assume that the circumbinary disk is co-planar with the SBHB orbit and co-rotating with it and therefore $\theta_3\equiv 0^\circ$,  $\phi_3 \equiv \phi$ and $i_3 \equiv i$ (i.e., the circumbinary disk frame is coincident with the SBHB frame). These relationships allow us to define the emissivity of surface elements in each of the three disks in the system. 

\subsection{Calculation of disk emissivities}

We assign emissivities $\epsilon_1$,  $\epsilon_2$ and $\epsilon_3$ to the primary, secondary and circumbinary disk, respectively. The emissivity of a gas element located at $(\xi_1,\varphi_1)$ in the mini-disk of the primary SBH can be expressed as a sum of  components due to the illumination by its own AGN ($\epsilon_{11}$) and the AGN associated with the secondary SBH ($\epsilon_{12}$)
\begin{equation}
\label{eq:emiss1}
\epsilon_1(\xi_1,\varphi_1)=
\epsilon_{11} + \epsilon_{12}= 
\epsilon_{10}\frac{h_1}{\xi_1^3} + 
\epsilon_{20}\frac{H \left(\cos i_1 \cos \theta_{M2}\right)\, \left|l_1 \left| \cos\theta_{M2} \right|-q\, h_2\right|}{\left[\xi_1^2 + l_1^2 - 2\xi_1 l_1 \left(\boldsymbol{\hat{\xi}_1} \cdot \boldsymbol{\hat{e}_r} \right) \right]^{3/2}}
\end{equation}
where $h_1$ and $h_2$ denote the sizes of the sources of continuum radiation associated with the two SBHs. As mentioned earlier, the subscripts indicate that $h_1$ and $h_2$ are dimensionless quantities in units of $M_1$ and $M_2$, respectively, and $l_1$ is the separation of the two AGN in units of $M_1$. Motivated by the X-ray studies of the broad iron line reverberation \citep{uttley14}, we assume that the sources of continuum radiation are compact and have spatial extents of $h_1=10$ and $h_2= 10$. Note that the term $q\, h_2$ in equation~\ref{eq:emiss1} represents conversion of $h_2$ into the units of $M_1$, for consistency with the rest of the properties calculated in the frame of the primary mini-disk, where $q=M_2/M_1$ is the SBH mass ratio. 

The second term of equation~\ref{eq:emiss1} captures the effect of the off-center illumination of the primary BLR by the secondary AGN. Figure~\ref{fig:single} illustrates that this effect can be seen by a distant observer only when the side of the mini-disk illuminated by the off-center AGN is facing the observer. Alternatively, whenever the secondary AGN is blocked by the primary mini-disk (from the observer's point of view) this effect will be absent. We describe these outcomes with the Heaviside step function, $H(\cos i_1 \cos \theta_{M2})$, which takes value ``1" whenever the secondary AGN belongs to the same half plane with the observer ($\cos i_1 \cos \theta_{M2}\geq 0$) and value ``0" otherwise. 

As discussed in Section~\ref{sec:appendixsingle}, the emissivity constants $\epsilon_{10}$ and $\epsilon_{20}$ are directly proportional to the luminosity of the two AGNs, which we assume are powered by accretion onto the SBHs. Therefore, we express the ratio of the two  constants as $\dot{m} = \epsilon_{20}/\epsilon_{10} = \dot{M_2}/\dot{M_1}$. Assuming relative scaling such that $\epsilon_{10} = 1$ implies $\epsilon_{20} = \dot{m}$ and equation~\ref{eq:emiss1} becomes
\begin{align}\label{eq:emiss2}
\begin{split}
\epsilon_1(\xi_1,\varphi_1)=\frac{10}{\xi_1^3} + \dot{m}\frac{H(\cos i_1 \cos \theta_{M2})\; |l_1|\cos\theta_{M2}|-10q |}{\left[\xi_1^2 + l_1^2 - 2\xi_1 l_1 \sin\theta_{M2}\left(\cos\varphi_1\cos\phi_{M2}+\sin\varphi_1\sin\phi_{M2}\right) \right]^{3/2}}
\end{split}
\end{align}
The vertical brackets in equations~\ref{eq:emiss1} and \ref{eq:emiss2} denote absolute values of the relevant quantities. Similarly, the emissivity of the secondary mini-disk can be expressed as:
\begin{equation}\label{eq:emiss3}
\epsilon_2(\xi_2,\varphi_2)=
\dot{m}\,\frac{10}{\xi_2^3} + \frac{H(\cos i_2 \cos \theta_{M1})\, 
\left| l_2 \left| \cos\theta_{M1}\right| -10/q\right|}
{\left[\xi_2^2 + l_2^2 + 2\xi_2 l_2 \sin\theta_{M1}\left(\cos\varphi_2\cos\phi_{M1}+\sin\varphi_2\sin\phi_{M1}\right) \right]^{3/2}}
\end{equation}

In the case of the circumbinary disk the illumination by both AGNs is off-center. We express its emissivity as that of a single accretion disk which center resides at the center of mass of the binary
\begin{equation}\label{eq:emiss4}
\epsilon_3(\xi_3,\varphi_3)=
\frac{10}{1+q}\frac{1}{(\xi_3^2+l_{31}^2 + 2\,\xi_3 \,l_{31}\cos \varphi_3)^{3/2}}+
\dot{m}\, \frac{10\,q}{1+q}\frac{1}{(\xi_3^2 + l_{32}^2 - 2\,\xi_3\, l_{32} \cos \varphi_3)^{3/2}}
\end{equation}
Because the coordinate system associated with the circumbinary disk coincides with the reference frame of the binary, all distances in equation~\ref{eq:emiss4} are normalized by the total mass of the binary $M$. Therefore, $\xi_3=r_3/M$ is the dimensionless radial distance of the gas element to the SBHB center of mass and $l_{31}= q/(1+q) (l/M)$ and $l_{32}= 1/(1+q) (l/M)$ are the dimensionless distances from the center of mass to the primary and secondary SBHs, respectively.

\subsection{Total flux of the composite emission line profile}

With known emissivities the flux from each disk in the SBHB system can be calculated as an integral over the surface area, according to equation~\ref{eq:flux1}. Before summing the fluxes to calculate the composite emission line profile we need to account for the Doppler shift of the photons emitted by the primary and secondary mini-disks due to their orbital motion. Because the circumbinary disk is at rest with respect to the SBHB center of mass we apply no shift to its emission. The emitted composite profile is therefore calculated in the reference frame of the binary.

Because the orbital velocities of SBHBs considered in this work are non-relativistic, the Doppler shifts associated with the orbital motion can be evaluated in classical limit. We therefore neglect the effect of relativistic boosting in this case as well as the gravitational redshift and lensing of photons that may arise in configurations when the two SBHs are in conjunction (i.e., lined up along the observer's line of sight). In classical limit, the Doppler shift associated with the orbital motion of the secondary mini-disk is $D_{\rm orb2}=1/\left(1-\boldsymbol{v_{\rm orb2}} \cdot \boldsymbol{ \hat{s}}\right)$ and 
\begin{equation}
\frac{\Delta \nu_{\rm orb2}}{\nu_0}=
\frac{\nu_{\rm obs}-\nu_0}{\nu_0}=
\frac{1}{1-\vec{v_{\rm orb2}}\cdot \vec{\hat{s}}} - 1 \simeq 
\vec{v_{\rm orb2}}\cdot \vec{\hat{s}}
\end{equation}
Here $\boldsymbol{v_{\rm orb2}}$ is the velocity vector of the secondary SBH measured in the frame of the binary in units of $c$ and  $\boldsymbol{\hat{s}}$ describes the orientation of the observer in the frame of the SBHB as defined earlier. Let $\boldsymbol{v_{\rm orb}}=\boldsymbol{v_{\rm orb2}}-\boldsymbol{v_{\rm orb1}}$ be the relative velocity vector of the two SBHs, as in Figure~\ref{fig:binary}, and
\begin{eqnarray}
&&v_{\rm orb}^2=\left(\frac{2}{l}-\frac{1}{a}\right)\\
&&\boldsymbol{v_{\rm orb}}=\dot{\boldsymbol{l}}=
v_r \,\boldsymbol{\hat{e}_r}+v_f\,\boldsymbol{\hat{e}_{\perp}}=\dot{l}\,\boldsymbol{\hat{e}_r}+l\dot{f}\,\boldsymbol{\hat{e}_{\perp}}\\
&&\boldsymbol{v_{\rm orb}}=\left[\frac{1}{a(1-e^2)}\right]^{1/2} \left[-e \sin f \,\boldsymbol{\hat{e}_r} + (1- e \cos f) \, \boldsymbol{\hat{e}_{\perp}} \right]
\end{eqnarray}
where $l$ and $a$ are in units of $M$ and $\boldsymbol{\hat{e}_r}=(\cos  f \,\boldsymbol{\hat{x}} + \sin f \,\boldsymbol{\hat{y}})$ and $\boldsymbol{\hat{e}_{\perp}}=(-\sin f\, \boldsymbol{\hat{x}} + \cos f \,\boldsymbol{\hat{y}})$ are the unit vectors parallel and perpendicular to $\boldsymbol{l}$, respectively. By conservation of momentum, $\vec{v_{\rm orb2}} = \vec{v_{\rm orb}} /(1+q) $, yielding the Doppler shifts for emission from the secondary and primary mini-disks
\begin{eqnarray}
\label{eq:orb1}
\Delta \nu_{\rm orb2}  =  
 \frac{\nu_0}{1+q}\left[\frac{1}{a(1-e^2)}\right]^{1/2}[- \sin f \sin i \cos\phi + (\cos f - e) \sin i \sin\phi] 
\end{eqnarray}
\begin{equation}
\label{eq:orb2}
\Delta \nu_{\rm orb1}= -q\, \Delta \nu_{\rm orb2}
\end{equation}

Because we are interested in the value of the total flux in some arbitrary normalized units (as opposed to the absolute units) the distance $d$ from the observer to the SBHB system can be omitted because it is the same for all three disks (see equation~\ref{eq:flux1}). Given the choice of dimensionless units employed in our calculation of emissivities in equations~\ref{eq:emiss2}--\ref{eq:emiss4}, the components of flux associated with the primary, secondary and circumbinary disk are proportional to $M_1^2$, $M_2^2$ and $M^2$, respectively. Therefore, in the expression for the total flux the relative contributions from each disk should be scaled in terms of the SBHB mass ratio $q$ as 
\begin{equation}
F_{\rm tot} = \frac{1}{(1+q)^2}\, F_1 + \frac{q^2}{(1+q)^2}\, F_2 + F_3
\end{equation}

\section{C: Dependence of Statistical Distribution Functions on $F_c$} \label{sec:appendixcutoff}

\begin{figure*}[b]
\centering
\includegraphics[width=.8\textwidth, clip=true]{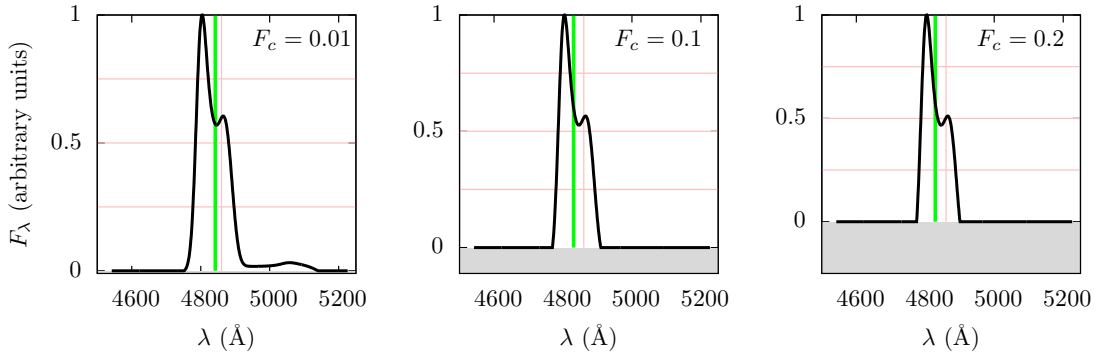}
\caption{Effect of different cutoff values $F_c$, representing some fiducial level of spectral noise, on a simulated line profile shape. For each profile the location of the centroid and the rest wavelength are marked by the green and pink vertical lines, respectively. The flux above the ``noise" level is rescaled so that the maximum flux measured at the peak wavelength has the value of 1.0.}
\label{fig:cutoff}
\end{figure*}

In this section we investigate the dependence of the distribution functions, characterizing the modeled profile shapes, on the value of $F_c$, a cutoff used to mimic some fiducial level of spectral noise. As noted in Section~\ref{sec:stats} we adopt $F_c = 0.01$ in calculation of statistical properties presented in this work but do not introduce actual fluctuations due to noise to the profiles. Figure~\ref{fig:cutoff} illustrates how different noise levels impact the line profile shapes, where in addition to $F_c = 0.01$ we examine the values of 0.1 and 0.2. With ``noise" subtracted from the profile, we rescale the flux above this cutoff so that the maximum flux measured at the peak wavelength always has the value of 1.0. 

One apparent consequence of the higher level of noise is that it can mask low intensity features present in the profile wings and hence, affect its statistical properties. The middle and right panel of Figure~\ref{fig:cutoff} show that when the noise conceals the low intensity feature between 4900 and $5200\AA$ the profile centroid (marked by the green vertical line) changes from $C=4842\AA$ to $4830\AA$ to $4828\AA$ for $F_c = 0.01$, 0.1 and 0.2, respectively. The sensitivity to the level of noise is particularly pronounced for higher order distribution functions that depend on the term $(\lambda_i - C)^n$, where index $n$ represents the order (see Section~\ref{sec:stats}). For example, the value of AI, which is proportional to the third moment, indicates that the profile changes from asymmetric (${\rm AI}=2.21$ at $F_c=0.01$) to relatively symmetric (${\rm AI}=0.35$ at $F_c=0.1$) with increasing $F_c$. Similarly, the value of KI, which is proportional to the fourth moment, indicates a transition from a cuspy (${\rm KI}=9.75$ at $F_c=0.01$) to a more boxy profile (${\rm KI}=2.02$ at $F_c=0.1$).

\begin{figure*}[t]
\centering
\includegraphics[width=.7\textwidth, clip=true]{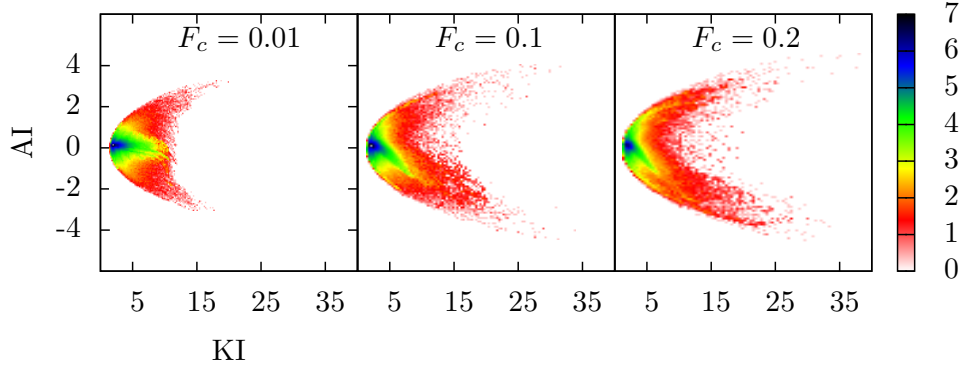}
\caption{AI-KI maps for profiles associated with eccentric SBHB systems calculated for different values of $F_c$. Map footprint and distribution of values vary for different adopted levels of noise indicating that AI and KI are sensitive functions of $F_c$.}
\label{fig:ai_noise}
\end{figure*}

This behavior of higher order distribution functions is illustrated in Figure~\ref{fig:ai_noise} which shows the AI-KI maps associated with eccentric SBHB systems and calculated for different values of $F_c$. They show that the overall footprint of the 2-dimensional distribution increases for the higher levels of noise while at the same time the average profile (traced by the blue and green colors) becomes more boxy.  

\begin{figure*}[t]
\centering
\includegraphics[width=.75\textwidth, clip=true]{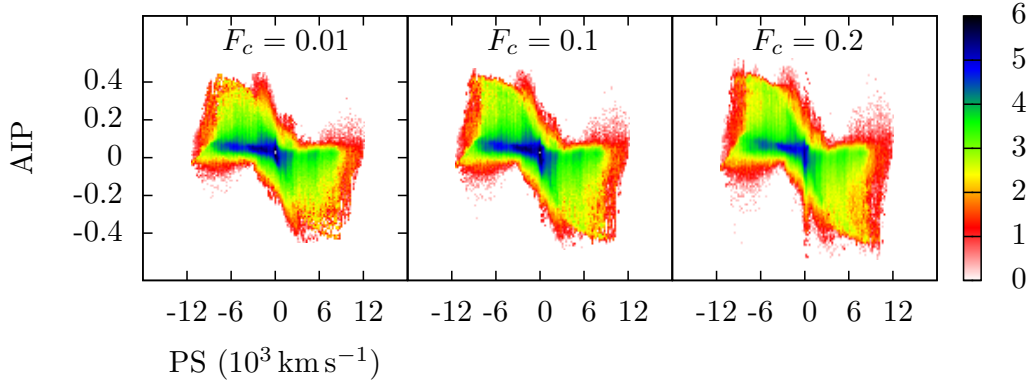}
\caption{AIP-PS maps for profiles associated with eccentric SBHB systems calculated for different values of $F_c$. Map footprint and distribution of values in different panels show little change as both AIP and PS are weak functions of $F_c$.}
\label{fig:aip_noise}
\end{figure*}

The value of the Pearson skewness coefficient, AIP, on the other hand exhibits a weak dependance on $F_c$.  For the profile in Figure~\ref{fig:cutoff} for example, ${\rm AIP}=0.21$ at $F_c=0.01$ and ${\rm AIP}=0.19$ at $F_c=0.1$. As discussed earlier, the AI and AIP provide different measures of the profile asymmetry. This is because the AI sensitively depends on the low intensity features in the profile wings and AIP diagnoses the asymmetry in the bulk of the profile. This property of AIP is captured in Figure~\ref{fig:aip_noise}, which shows the AIP-PS maps associated with eccentric SBHB systems. The map footprint and distribution of values in different panels show little change as both AIP and PS are  weak functions of $F_c$.

%%%%%%%%%%%%%%%%%%%%%%%%%%%%%%%%%%%%%%%%%%%%%%%%%%%%%
%%% REFERENCES
%%%%%%%%%%%%%%%%%%%%%%%%%%%%%%%%%%%%%%%%%%%%%%%%%%%%%
\bibliographystyle{apj}
\bibliography{apj-jour,smbh}

%%%%%%%%%%%%%%%%%%%%%%%%%%%%%%%%%%%%%%%%%%%%%%%%%%%%%
%%% FIGURES
%%%%%%%%%%%%%%%%%%%%%%%%%%%%%%%%%%%%%%%%%%%%%%%%%%%%%

%%%%%%%%%%%%%%%%%%%%%%%%%%%%%%%%%%%%%%%%%%%%%%%%%%%%%
%%% TABLES
%%%%%%%%%%%%%%%%%%%%%%%%%%%%%%%%%%%%%%%%%%%%%%%%%%%%%

\end{document}